\documentclass[aip,jcp,reprint]{revtex4-1}
\usepackage[dvips]{color}
\usepackage{amsmath}
\usepackage{amsfonts}
\usepackage{xfrac}
\usepackage{hyperref}
\usepackage{graphicx}
\usepackage{natbib}
\usepackage{epstopdf}
\usepackage{dcolumn}
\usepackage{subfigure}
\usepackage{lastpage}
\usepackage{fancyhdr}
\usepackage[titletoc]{appendix}
\usepackage{bm}

\begin{document}

\title{Dynamics and control of DNA sequence amplification}
\author{Karthikeyan Marimuthu}
\affiliation{Department of Chemical Engineering and Center for Advanced Process Decision-Making, Carnegie Mellon University, Pittsburgh, PA 15213}
\author{Raj Chakrabarti}
\thanks{To whom correspondence should be addressed. Email:  raj@pmc-group.com; rajc@andrew.cmu.edu}
\affiliation{Department of Chemical Engineering and Center for Advanced Process Decision-Making, Carnegie Mellon University, Pittsburgh, PA 15213}
\affiliation{Division of Fundamental Research, PMC Advanced Technology, Mount Laurel, NJ 08054}
\date{\today}

\begin{abstract}
DNA amplification is the process of replication of a specified DNA sequence \emph{in vitro} through time-dependent manipulation of its external environment. A theoretical framework for determination of the optimal dynamic operating conditions of DNA amplification reactions, for any specified amplification objective, is presented based on first-principles biophysical modeling and control theory. Amplification of DNA is formulated as a problem in control theory with optimal solutions that can differ considerably from strategies typically used in practice. Using the Polymerase Chain Reaction (PCR) as an example, sequence-dependent biophysical models for DNA amplification are cast as control systems, wherein the dynamics of the reaction are controlled by a manipulated input variable. Using these control systems, we demonstrate that there exists an optimal temperature cycling strategy for geometric amplification of any DNA sequence and formulate optimal control problems that can be used to derive the optimal temperature profile. Strategies for the optimal synthesis of the DNA amplification control trajectory are proposed. Analogous methods can be used to formulate control problems for more advanced amplification objectives corresponding to the design of new types of DNA amplification reactions.
\end{abstract}
\maketitle
\section{Introduction} \label{introduction}
DNA amplification refers to a class of methods that can achieve geometric growth of the number of double-stranded DNA (dsDNA) molecules through repeated polymerization on single-stranded DNA (ssDNA) templates. Such methods have arguably become the central technology of experimental molecular biology and biochemistry, due to the fact that DNA amplification is required almost universally in applications ranging from molecular cloning to DNA sequencing. The most common DNA amplification reaction is the polymerase chain reaction (PCR), a cyclic amplification process that can produce millions of copies of dsDNA molecules starting from a single molecule, by repeating three steps -- (i) dsDNA denaturation, (ii) oligonucleotide primer annealing to the resulting ssDNAs, and (iii) polymerase-mediated extension steps to produce two dsDNA molecules -- resulting in geometric growth of the number of DNA molecules.

In addition to its importance as the primary method for preparing copies of dsDNA, DNA amplification is a model system for the study of fundamental properties of the dynamics and control of sequence replication \cite{retkute2011dynamics,retkute2012mathematical,sharma2012error,zhang2006reconstructing}.
DNA amplification is concerned with the promotion of replication of a given sequence by time-dependent manipulation of the external environment. In this regard it is important to distinguish the theory of nucleic acid amplification dynamics from that of Darwinian evolution. Darwinian evolutionary theory is concerned with the 
problem of sequence evolution or optimization through replication in the presence of a static or specified time-varying external environment. This encompasses strategies such as laboratory directed evolution, which apply external selection pressure to a population of sequences in order to promote the optimization of fitness through sequence evolution \cite{chakrabarti2005computational}.
Darwinian evolution assumes the ability of a system to both self-replicate and mutate. It is generally acknowledged that during the early stages of prebiotic evolution, protocells consisting of compartmentalized nucleic acids were not capable of self-replication under time-invariant environmental conditions.
As such, they required time-varying environmental inputs to improve replication efficiency \cite{Szostak2008,Szostak2009} prior to the onset of Darwinian evolution. For example, time-varying temperature profiles capable of driving nucleic acid replication may have occurred on the early earth  due to diurnal cycles of heating and cooling or temperature gradients
generated by geothermal activity \cite{Szostak2009}. Mutations occurring during these cycles of environmentally controlled replication may then have led to functional nucleic acids that ultimately developed the ability to self-replicate \cite{Szostak2012}.
Thus, unlike the Darwinian paradigm of selective replication of the fittest sequences in an uncontrolled environment, control of DNA amplification may involve selection of time-varying inputs that promote the survival and replication of otherwise less fit sequences. Hence the theory of DNA amplification dynamics is distinct from the theory of evolutionary dynamics \cite{chakrabarti2008mutagenic, kloster2004simulation} -- although they are closely related since both are concerned with the dynamics of sequence replication.

In DNA amplification by the polymerase chain reaction, the time-varying environmental input is achieved through thermal cycling of the reaction mixture.
In order to compute the optimal temperature cycling protocol for the reaction - which is sequence-specific - a dynamic model for DNA amplification capable of predicting the evolution of reaction products for general sequences and operating conditions is required. However, in the absence of such a dynamic model, the operating conditions for PCR reactions are typically selected based on analysis of reaction thermodynamics, and, to a lesser extent, qualitative analysis of reaction kinetics. Reductions in cycle efficiency (either through decreased reaction yield or specificity compared to the theoretical maximum values) therefore commonly occur - and due to geometric growth, this can result in dramatically diminished efficiency of the overall reaction. This has led to considerable interest in the development of methods for improvement of DNA amplification efficiency \cite{Chakra1,Chakra2}.

A general approach to dynamic optimization of DNA amplification could be used to systematically design new types of amplification reactions \cite{li2008replacing} corresponding to specified sequence amplification objectives, as well as to enhance existing reactions.
A simple example of a temperature cycling protocol that involves modifications to the conventional prescription is the use of two-step PCR cycles, wherein annealing and extension occur simultaneously at a properly chosen temperature. Another important example of modified temperature cycling is the method of co-amplification at reduced denaturation temperature (COLD-PCR) \cite{li2008replacing}, which enriches the amplification of low copy number mutant sequences that are associated with early cancer prognosis, in the presence of a vastly more concentrated wild-type DNA background. This approach introduces intermediate steps in each PCR cycle to achieve mutant enrichment. Still another example arises in the problem of multiplex PCR, where annealing temperatures must be chosen such that several primers can simultaneously anneal, while avoiding the formation of mismatched hybrids. Over the past two decades, many other variants of DNA amplification have been invented based on the notions of DNA denaturation, annealing and polymerization, each tailored to a particular amplification objective. Each such reaction (which is typically assigned its own acronym) is based on a temperature cycling protocol determined through analysis of reaction thermodynamics and kinetics. Formulation of DNA amplification problems as optimal control problems should provide a general framework for  the discovery of new classes of such reactions.

This paper is concerned with the establishment of a foundation for the optimal control of DNA amplification reactions, which can be used for the automated computation (rather than qualitative selection) of temperature cycling protocols. Control theory provides a convenient means of parameterizing the interaction of a reaction network for molecular sequence replication with an external environment. To our knowledge, a theory of optimal environmental control of molecular sequence replication has not been proposed. We formulate this theory for the model system of DNA amplification via the polymerase chain reaction.

For dynamic optimization of DNA amplification, a sequence-dependent state space model is required. A sequence-dependent state space model is a system of differential equations that, when solved, describe the dynamics of the evolution of particular biopolymer sequences, along with algebraic constraints and specified parameters (e.g., rate parameters such as activation energies and pre exponential factors) whose values are either predicted based on first-principles theory, independently measured in offline experiments, or indirectly estimated through online measurement of observable quantities (such as total DNA concentration).

In this paper, we use first-principles sequence-dependent kinetic models recently introduced in \cite{biophys2014} and \cite{Karthik1} to formulate amplification of DNA as a problem in control theory with optimal solutions that can differ considerably from strategies typically used in practice. First, the notion of sequence-dependent control systems of biochemical reaction networks is introduced. Control systems corresponding to several DNA amplification models are then formulated and compared. Next, we formulate optimal control problems based on the sequence-dependent kinetic model for DNA amplification and specified objectives. Besides sequence dependence, another novel feature of control of DNA amplification is the cyclic nature of optimal temperature control strategies, which are usually assumed to be periodic. We show that the optimal control strategy for DNA amplification will change depending on the stage of the reaction, due to changes in the availability of resources required for replication, and demonstrate how the control problem can be formulated to enable prediction of aperiodic manipulated input functions. Finally, strategies for the optimal synthesis are proposed for each stage of amplification.

\section{Sequence and Temperature dependent Kinetic Model for DNA Amplification} \label{model_develop}
In \cite{biophys2014},
a sequence and temperature dependent state space model for DNA amplification was developed by the authors, and model parameters were estimated using experimental datasets. In this section we summarize and further develop this model, which is used in the present work to study the control of DNA amplification.
\subsection{Annealing}
Reaction \ref{R1} represents the annealing reaction of an oligonucleotide primer (P) and single stranded template DNA (S):
\begin{align}
S+P \underset{k_{r}}{\overset {k_{f}}{\rightleftharpoons}} SP  \tag{$R_{1}$} \label{R1}
\end{align}
Marimuthu and Chakrabarti \cite{Karthik1} developed a theoretical framework that couples the equilibrium thermodynamics and relaxation kinetics to estimate the sequence and temperature dependent annealing rate constants. 
The forward and reverse rate constants $k_f,k_r$ of annealing reaction \ref{R1} are expressed in terms of the equilibrium constant $K_{annealing}$ and relaxation time $\tau$, where
\begin{equation}
K_{annealing} = exp\left(-\Delta G/RT\right) = k_f/k_r   \label{eq_relation} \\
\end{equation}
and
\begin{equation}
\tau = \frac{1}{k_f\left([S_{eq}] + [P_{eq}]\right) + k_r}.   \label{tau_reln}
\end{equation}
$\Delta G$ in Eq. (\ref{eq_relation}) can be estimated using the Nearest Neighbor Model of hybridization thermodynamics \cite{SantaLucia,Hicks}.
$\tau$, which is a characteristic time constant that determines the evolution of reaction coordinates after perturbation from equilibrium, can be computed at a chosen temperature either one- or  two-sided melting master equations \cite{Karthik1}.  
For one-sided melting, the master equation is that of a biased one-dimensional random walk with partially reflecting boundary conditions. The walk is biased because the forward and reverse reaction rates are not equal. The master equation for one-sided melting of a homogeneous sequence is given by
\begin{align}
\frac{\partial}{\partial t} \rho_{a}\left(0,t\right) &= - k^a_1 \rho_{a}\left(0,t\right) + k^a_{-1} \rho_{a}\left(1,t\right)  \nonumber \\
\frac{\partial}{\partial t} \rho_{a}\left(j,t\right) &= - \left(k^a_1 + k^a_{-1}\right) \rho_{a}\left(j,t\right) + k^a_1  \rho_{a}\left(j-1,t\right) +\nonumber \\
&~~~~~~~+k^a_{-1}  \rho_{a}\left(j+1,t\right),~~~~~~j=1,\cdots,n-1  \nonumber \\
\frac{\partial}{\partial t}  \rho_{a}\left(n,t\right) &= -k^a_{-1}\rho_{a} \left(n,t\right) + k^a_1 \rho_{a}\left(n-1,t\right)
\label{master_eqn_hyb}
\end{align}
where $\rho_{a} \left(j,t\right)$ denotes the occupation probability of duplexes with $i$ base pairs in the annealed/hybridized state at time $t$. The transition rates $k^a_{1}$ and $k^a_{-1}$ denote the forward and reverse rate constants for hybridization of a base pair. 
Eigendecomposition of the corresponding state space matrix that is equivalent to the master equation represented by Eq. \ref{master_eqn_hyb} directly provides the relaxation time as the negative reciprocal of the maximum eigenvalue of the state space matrix.
Marimuthu and Chakrabarti \cite{Karthik1} developed a theoretical method to calculate the sequence and temperature dependent relaxation time of reaction \ref{R1} and provided correlations to estimate model parameters $k^a_{1}$,  $k^a_{-1}$, and $\sigma$. They also generalized this approach to accommodate heterogeneous sequences and two-sided oligonucleotide melting models.

Thus, it is possible to calculate the left hand side of Eq. (\ref{tau_reln}) for any oligonucleotide sequence. Then, solving Eq. (\ref{eq_relation}) and Eq. (\ref{tau_reln}) simultaneously, we can obtain the rate constants $k_f$ and $k_r$. 

\subsection{Polymerase Binding and Extension}
Marimuthu \textit{et al} \cite{biophys2014} 
estimated the enzyme binding and extension rate constants experimentally using the theory of processivity and Michaelis-Menten (MM) kinetics. Reaction \ref{R2} represents the simplified mechanism of these reactions.
\begin{align}
E + D_i  &\underset{k_{-1}^e}{\overset {k_1^e}{\rightleftharpoons}} E.D_i  + N  \underset{k_{-2}^e}{\overset {k_2^e}{\rightleftharpoons}} E.D_{i} .N {\overset {k_{cat}}{\rightarrow}} ED_{i+1} \underset{k_{1}^e}{\overset {k_{-1}^e}{\rightleftharpoons}} E + D_{i+1},\nonumber  \\ 
&i=0,\cdots,n-1. \tag{$R_2$} \label{R2}
\end{align}
Here $D_i$ denotes a partially extended primer-template DNA with $i$ added bases ($D_0\equiv SP$), $E.Di$ denotes its complex with polymerase enzyme $E$ and $N$ denotes nucleotide. For the purpose of model parameter estimation, the reaction mechanism \ref{R2} is studied under so-called single hit conditions in which polymerase concentrations are sufficiently low that the probability of enzyme re-association is approximately 0. 
Hence enzyme-template association occurs only during the initial equilibration of enzyme with SP. Assuming the intermediate $E.D_i.N$ is at steady state during the initial rate measurements, it can be omitted from the model and the transition rate for nucleotide addition is $\frac{k_{cat}}{K_N}[N]\equiv k^e$, where $K_N = (k_{-2}^e + k_{cat})/k_2^e$. In a single molecule continuous-time Markov chain formulation, evolution of the $E.D_i$ and $D_i$ molecules can then be written in terms of the probabilities of states $\left[E.D_0,D_0, E.D_1,D_1,...\right]$ 
according to the master equation 
\begin{align}
&\frac{\partial}{\partial t} \rho_e \left(0,t\right) = -\left(k^e + k_{-1}^e\right)  \rho_e\left(0,t\right)  \nonumber \\
&\frac{\partial}{\partial t} \rho_e\left(j,t\right) = k_{-1}^e  \rho_e\left(j-1,t\right),~~j=1,3,5,\cdots,2n-3 \nonumber \\
&\frac{\partial}{\partial t} \rho_e\left(j+1,t\right) = - (k^e+k_{-1}^e)\rho_e(j+1,t) + k^e  \rho_e\left(j-1,t\right)  \nonumber \\
&\frac{\partial}{\partial t} \rho_e\left(2n-1,t\right) = k_{-1}^e  \rho_e\left(2n-2,t\right);~~~~~~\rho_e\left(0,0\right)  = 1 \nonumber \\
\label{master extension}
\end{align}
where $\rho_e\left(j,t\right)$ denotes the probability of the polymerase being in state $j$ at time $t$, $k_{-1}^e ~dt$ denotes the conditional probability of transition $E.D_i \rightarrow E + D_i$  in time $dt$, and $k ~dt$ denotes the conditional probability of transition $E.D_i \rightarrow E.D_{i+1}$ in time $dt$. 
The solution to the master equation can be obtained analytically through Jordan decomposition of the degenerate fundamental matrix (or Laplace transformation) of the following equivalent state space representation of (\ref{master extension}) with rescaled time ($t' = (k^e+k_{-1}^e)t$):
\begin{equation}
\frac{d\rho }{dt'}=\left[
\begin{array}{ccccccc}
-1 & 0 & 0 & 0 & 0  & \cdots & 0\\
\frac{k_{-1}^e}{k_{-1}^e+k^e}& 0 & 0 & 0 & 0 & \cdots & 0\\%
\frac{k^e}{k_{-1}^e+k^e} & 0 & -1 & 0 & 0 & \cdots & 0\\
0& 0 & \frac{k_{-1}^e}{k_{-1}^e+k^e} & 0 & 0 &  \cdots & 0\\%
0& 0 & \frac{k^e}{k_{-1}^e+k^e} & 0 & -1 &  \cdots & 0\\%
\vdots  & \vdots & \vdots & \vdots & \vdots   & \ddots & \vdots \\
0& 0 & 0 & 0 & 0 & \cdots & 0%
\end{array}
\right]\rho(t')
\label{eigen_ss_binding}
\end{equation}
where
$\rho(t') = [p_{E.D_0}(t'),p_{D_0}(t'),...]^T$ is the probability distribution of states at time $t'$.

For the specified boundary conditions, the system converges to the distribution
\begin{equation}
\pi = \left[0,  \frac{k_{-1}^e}{k^e+k_{-1}^e}, 0, \frac{k^ek_{-1}^e}{\left(k^e+k_{-1}^e\right)^2},..., 1 - \sum_{i=1}^{n-1} \frac{(k^e)^{i-1}k_{-1}^e}{\left(k^e+k_{-1}^e\right)^{i}} \right]^T
\end{equation}
Now let $p$ denote the conditional probability of the polymerase \emph{not} dissociating at position $i$. The probability of dissociation at position/time $i$ is
\begin{equation}
p_{off}(i) = \left(1-p\right) p^{i-1}  \label{process_defn}
\end{equation}
which describes the stationary distribution of $D_i$ molecules at the end of the experiment.
Equating \ref{process_defn} with the corresponding components of $\pi$ gives  
\begin{align}
k_{-1}^e = \frac{k_{cat}}{K_N}\left[N\right]\frac{1-p}{p}\label{process}
\end{align}
Eq. (\ref{process}) expresses the enzyme dissociation rate constant in terms of the microscopic processivity parameter $p$ and MM kinetic parameters $k_{cat}$ and $K_N$. Typically, polymerase processivity is characterized in terms of $\mathrm{E}[i_{off}]=\sum_i ip_{off}(i)\approx\frac{1}{1-p}$.
$k_{cat}$ and $K_N$ are obtained from Michaelis-Menten measurements of the initial rate of polymerase-mediated extension \cite{biophys2014}.
With the value of $k_{-1}$ estimated above and the temperature-dependent equilibrium constant for polymerase binding \cite{Datta}, the temperature-dependent polymerase association rate constant $k_1$ can also be estimated. 

\subsection{Long DNA Melting}
In the present study the DNA melting reaction is assumed to be 100\% efficient. 
However, Marimuthu \textit{et al}. \cite{biophys2014} presented a theoretical framework for the estimation of the DNA melting rate constants using the concepts of domain melting and relaxation kinetics.
The Poland-Scheraga algorithm \cite{garel2004,richard2004poland}or MELTSIM \cite{Blake} can be used to identify the discrete domains of a long DNA molecule, and the overall equilibrium constant for the melting of each domain can be obtained via the following equation:
\begin{equation}
K_{loop} =\sigma_c f(N) \prod_{i=1}^N s_i \label{domain_melting1}
\end{equation}
where 
$\sigma_c$ is referred as a cooperativity parameter \cite{Blake} that constitutes a penalty on the statistical weight for melting of the domain ($\prod_{i=1}^N s_i$) due to the free energy cost of dissociation of an internal base pair and $f(N)$ is a loop closure function that introduces a length dependence to the free energy cost.  A method for calculation of $\sigma_c$ and $f(N)$ for a given domain has been explained in \cite{Blake}. All of these domains melt based on the two-state theory described in the annealing model section, and therefore the relaxation time for melting of each domain can be obtained analogously. The domain that possesses the largest relaxation time represents the rate limiting domain and this relaxation time is the relaxation time of the overall melting process. Solving the corresponding equations (\ref{eq_relation}) and (\ref{tau_reln}) for long DNA melting, the rate constants for the melting reaction can be estimated.

\section{Dynamic models and control systems for DNA amplification}
In this section we establish a control theoretic framework for dynamic optimization of DNA amplification.  Several classes of control systems, differing in their representations of the sequence and temperature dependence of chemical kinetic rate constants, are introduced. The most general control system is based on the sequence- and temperature-dependent state space model for DNA amplification described above. Other control systems are based on simplified PCR models previously proposed in the literature \cite{Mehra,Hsu,Gevertz,Stolovitzky} and correspond to approximations to the model described in Section \ref{model_develop}.
For these latter models, we only consider the previously developed model structures with rate constants estimated based on the above approach, since Marimuthu \textit{et al} \cite{biophys2014} have shown that the rate constants that were used in the previously proposed models are thermodynamically inconsistent.

Denoting by $u_i$ the $i^{th}$ rate constant, which can be manipulated as a function of time, and by $x$ the vector of species concentrations, a general state space model for chemical reaction kinetics can be written as follows:

\begin{equation}
\frac{dx}{dt} = f(x,u) = g_0\left(x\right) + \sum_{i = 1}^m u_ig_i\left(x\right),  \hspace{2mm} x(0) = x_0 \label{seq_gform}
\end{equation}
where $$x \in  \mathbb{R}^{n}, u \in  \mathbb{R}^{m}, g_i\left(x\right) :  \mathbb{R}^{n}\rightarrow \mathbb{R}^{n}$$

The above form is called a \textit{control-affine system} and $g_0\left(x\right)$ is called the drift vector field. The $g_i$\rq{}s associated with $u_i$\rq{}s are referred to as control vector fields. The notation $u_i$ is conventionally used to denote  manipulated input variables in control theory. However, in what follows we will use the notation $k_i$ since we will be dealing exclusively with chemical rate constants.

For PCR amplification reactions, a general dynamic model (assuming all $g_i$\rq{}s are control vector fields coupled to time-varying rate constants) can be written
\begin{equation}
\frac{dx(t)}{dt} = \sum_{i=1}^{10} k_i g_i\left(x\left(t\right)\right)  \label{steq_gi}
\end{equation}
$$k \in \mathbb{R}^{10};~k \geq 0, ~~x,g_i\left(x\right) \in \mathbb{R}^{4n+9}, ~x \geq 0,~~h\left(x\right) \in  \mathbb{R}^{5},~h\left(x\right) =0$$
$$ k = \left[k_m \hspace{2mm} k_{-m} \hspace{2mm} k_{f}^1 \hspace{2mm} k_{r}^1 \hspace{2mm} k_{f}^2 \hspace{2mm} k_{r}^2 \hspace{2mm}  k_1^e \hspace{2mm} k_{-1}^{e} \hspace{2mm} \frac{k_{cat}}{K_N} \hspace{2mm} k\rq{}_{cat} \right]$$
where $n$ denotes the number of base pairs, $g_1\left(x\right)$ and $g_2\left(x\right)$ represent the melting reaction alone, $g_3\left(x\right)$ to $g_6\left(x\right)$ represent the annealing reactions alone,  $g_7\left(x\right)$ to $g_{10}\left(x\right)$ represent the extension reactions and $h\left(x\right)$ denotes a set of 5 independent nonlinear constraints enforcing chemical mass balance, which cause the system to evolve on a state manifold $X \subset \mathbb{R}^{4n+9}$  that is of dimension $4n+4$. Hence the dimension of the state manifold is also sequence-dependent.  $x$, $g\left(x\right)$ and $h\left(x\right)$ for a simplex PCR with $n = 2$ are presented below. 
\begin{widetext}
$$x = \left[S_1, S_1P_1, E.S_1P_1, D_1^1, E.D_1^1, E.D_2^1, S_2, S_2P_2, E.S_2P_2, D_1^2, E.D_1^2, E.D_2^2, DNA, P_1,  P_2, N, E \right]^T$$
\begin{align*}
g_1(x) &= \left[x_{13} \hspace{2mm} 0 \hspace{2mm} 0 \hspace{2mm} 0 \hspace{2mm} 0 \hspace{2mm} 0 \hspace{2mm} x_{13} \hspace{2mm} 0 \hspace{2mm} 0 \hspace{2mm} 0 \hspace{2mm} 0 \hspace{2mm} 0 \hspace{2mm} -x_{13} \hspace{2mm} 0 \hspace{2mm} 0 \hspace{2mm} 0 \hspace{2mm} 0 \right]^T \\
g_2(x) &= \left[-x_1x_7 \hspace{2mm} 0 \hspace{2mm} 0 \hspace{2mm} 0 \hspace{2mm} 0 \hspace{2mm} 0 \hspace{2mm} -x_1x_7 \hspace{2mm}  \hspace{2mm} 0 \hspace{2mm} 0 \hspace{2mm} 0 \hspace{2mm} 0 \hspace{2mm} x_1x_7 \hspace{2mm} 0 \hspace{2mm} 0 \hspace{2mm} 0 \hspace{2mm} 0 \right]^T \\
g_3(x) &= \left[-x_1x_{14} \hspace{2mm} x_1x_{14} \hspace{2mm} 0 \hspace{2mm} 0 \hspace{2mm} 0 \hspace{2mm} 0 \hspace{2mm} 0 \hspace{2mm} 0 \hspace{2mm} 0 \hspace{2mm} 0 \hspace{2mm} 0 \hspace{2mm} 0 \hspace{2mm} 0 \hspace{2mm} -x_1x_{14} \hspace{2mm} 0\hspace{2mm} 0 \hspace{2mm} 0 \right]^T \\
g_4(x) &= \left[x_2 \hspace{2mm}  -x_2 \hspace{2mm} 0  \hspace{2mm} 0 \hspace{2mm} 0 \hspace{2mm} 0 \hspace{2mm} 0 \hspace{2mm} 0 \hspace{2mm} 0 \hspace{2mm} 0 \hspace{2mm} 0 \hspace{2mm} 0 \hspace{2mm} 0 \hspace{2mm}  x_2 \hspace{2mm} 0\hspace{2mm} 0 \hspace{2mm} 0 \right]^T \\
g_5(x) &= \left[0 \hspace{2mm} 0 \hspace{2mm} 0 \hspace{2mm} 0 \hspace{2mm} 0 \hspace{2mm} 0 \hspace{2mm}  -x_7x_{15} \hspace{2mm} x_7x_{15} \hspace{2mm}  0 \hspace{2mm}  0 \hspace{2mm} 0 \hspace{2mm} 0 \hspace{2mm} 0 \hspace{2mm} 0 \hspace{2mm} -x_7x_{15}\hspace{2mm} 0 \hspace{2mm} 0 \right]^T \\
g_6(x) &= \left[0 \hspace{2mm} 0 \hspace{2mm} 0 \hspace{2mm} 0 \hspace{2mm} 0 \hspace{2mm} 0 \hspace{2mm}  x_{8} \hspace{2mm} -x_{8} \hspace{2mm} 0 \hspace{2mm} 0 \hspace{2mm} 0 \hspace{2mm} 0 \hspace{2mm} 0 \hspace{2mm} 0 \hspace{2mm} x_{8}\hspace{2mm} 0 \hspace{2mm} 0 \right]^T \\
g_7(x) &= x_{17}\left[0 \hspace{2mm}  -x_2 \hspace{2mm} x_2 \hspace{2mm} -x_4  \hspace{2mm} x_4 \hspace{2mm} 0 \hspace{2mm} 0 \hspace{2mm}  -x_{8} \hspace{2mm}  x_{8} \hspace{2mm}  -x_{10} \hspace{2mm}  x_{10} \hspace{2mm} 0 \hspace{2mm} 0 \hspace{2mm} 0 \hspace{2mm} 0\hspace{2mm} 0 \hspace{2mm} -\left(x_2 + x_4 + x_{8} + x_{10}\right) \right]^T \\
g_8(x) &= \left[0 \hspace{2mm} x_3 \hspace{2mm} -x_3 \hspace{2mm}  x_5 \hspace{2mm} -x_5 \hspace{2mm} 0 \hspace{2mm} 0 \hspace{2mm}  x_{9} \hspace{2mm}  -x_{9} \hspace{2mm}  x_{11} \hspace{2mm}  -x_{11} \hspace{2mm}0 \hspace{2mm} 0 \hspace{2mm} 0 \hspace{2mm} 0\hspace{2mm} 0 \hspace{2mm} \left(x_3 + x_5 + x_{9} + x_{11}\right) \right]^T \\
g_9(x) &= x_{16}\left[0 \hspace{2mm} 0 \hspace{2mm} -x_3 \hspace{2mm} 0 \hspace{2mm}  x_3-x_5 \hspace{2mm} x_5 \hspace{2mm} 0 \hspace{2mm} 0  \hspace{2mm} -x_{9} \hspace{2mm} 0 \hspace{2mm}  x_{9}-x_{11} \hspace{2mm} x_{11} \hspace{2mm} 0 \hspace{2mm} 0\hspace{2mm} 0 \hspace{2mm}  \left(x_3 + x_5 + x_{9} + x_{11}\right) \hspace{2mm} 0 \right]^T \\
g_{10}(x) &= \left[0 \hspace{2mm} 0 \hspace{2mm} 0 \hspace{2mm} 0 \hspace{2mm} 0 \hspace{2mm}  -x_6 \hspace{2mm} 0 \hspace{2mm} 0 \hspace{2mm} 0 \hspace{2mm} 0 \hspace{2mm} 0 \hspace{2mm}  -x_{12} \hspace{2mm} 0 \hspace{2mm} 0 \hspace{2mm} 0 \hspace{2mm} 0 \hspace{2mm} x_6 + x_{12} \right]^T \\
\end{align*}
\end{widetext}

\noindent

\begin{align}
h_1\left(x\right) &= x_{14}^0 - x_1^0 + x_1 - x_{14} \\
h_2\left(x\right) &= x_{15}^0 - x_7^0 + x_7 - x_{15} \\
h_3\left(x\right) &= x_1^0 + x_7^0 - \left(\sum_{i=1}^{13} x_{i}\right) -\left(\frac{x_{16}}{K_N}\right) \left(\sum_{i=3,5,9,11} x_i\right)  \\
h_4\left(x\right) &= \frac{x_{16}^0 -\left(\sum_{i=4,5,10,11} x_i\right) - 2\left(x_6 + x_{12} + x_{13}\right) }{ \left(1 + \frac{1}{K_N}\right)\left(\sum_{i=5,11} x_i\right) } - x_{16} \\
h_5\left(x\right) &= x_{17}^0 - \left(1 + \frac{x_{16}}{K_N}\right)\left(\sum_{i=3,5,9,11} x_i\right) - x_6 - x_{12}
\end{align}

In the above definition of the components of the vector $x$, $DNA=D_n^1=D_n^2$.

Additional constraints, including equality constraints, may also be applied to the vector of rate constants $k$, due e.g. to their dependence on a single manipulated input variable, temperature.  As shown in \cite{biophys2014}, except for the extension reaction rate constants, all other rate constants are sequence-dependent. Therefore, the equality and inequality constraints on controls are sequence-dependent.  As a result of this the $k_ig_i$ in Eq. (\ref{steq_gi}) are also sequence-dependent.

We now introduce several types of control systems that are based on the above general formulation, but differ in terms of additional constraints they apply to the controls $k_i$ and the approximations they make regarding the sequence and temperature dependence of these controls.
\subsection{Staged Time Invariant (On-Off) DNA Amplification Model (STIM)}
Mehra and Hu \cite{Mehra} treated melting, annealing and extensions steps independently and did not consider the dissociation of $E.D_i$ molecules into $E$ and $D_i$ in reaction \ref{R2}. The state equations of the melting, annealing, and extension reactions were solved independently. Also, the rate constants of the respective reactions were held constant \cite{Mehra}.
We call this kind of PCR model a 'Staged' or 'On-Off' time invariant DNA amplification model, since the control vector fields are assumed to be either turned on or off at each step of a PCR cycle. Table \ref{model_comparison_table} presents its model structure. The system is controlled by manipulating the switching times between steps.

\subsection{Time Invariant DNA Amplification Model (TIM)}
Stolovitzky and Cecchi \cite{Stolovitzky} combined melting, annealing and extension together and formed a state space system for an overall reaction time (summation of the reaction times of melting, annealing, and extension). The rate constants are, however, held constant; i.e., $k$ is not a function of time. For example, annealing (primer hybridization) rate constants are not changed based on the annealing and extension temperatures. We call this kind of PCR model a time invariant DNA amplification model and its model structure is given in Table \ref{model_comparison_table}. The definitions of $x_i$, $k_i$ and $g_i$ are the same as those given in the staged time invariant DNA amplification model. Note that, unlike the STIM, the TIM system is formally not a control system since there are no manipulated input variables.
\subsection{Time Varying DNA Amplification Model with Drift (TVMD)}
\begin{figure} 
\includegraphics[width=8cm,height=6cm]{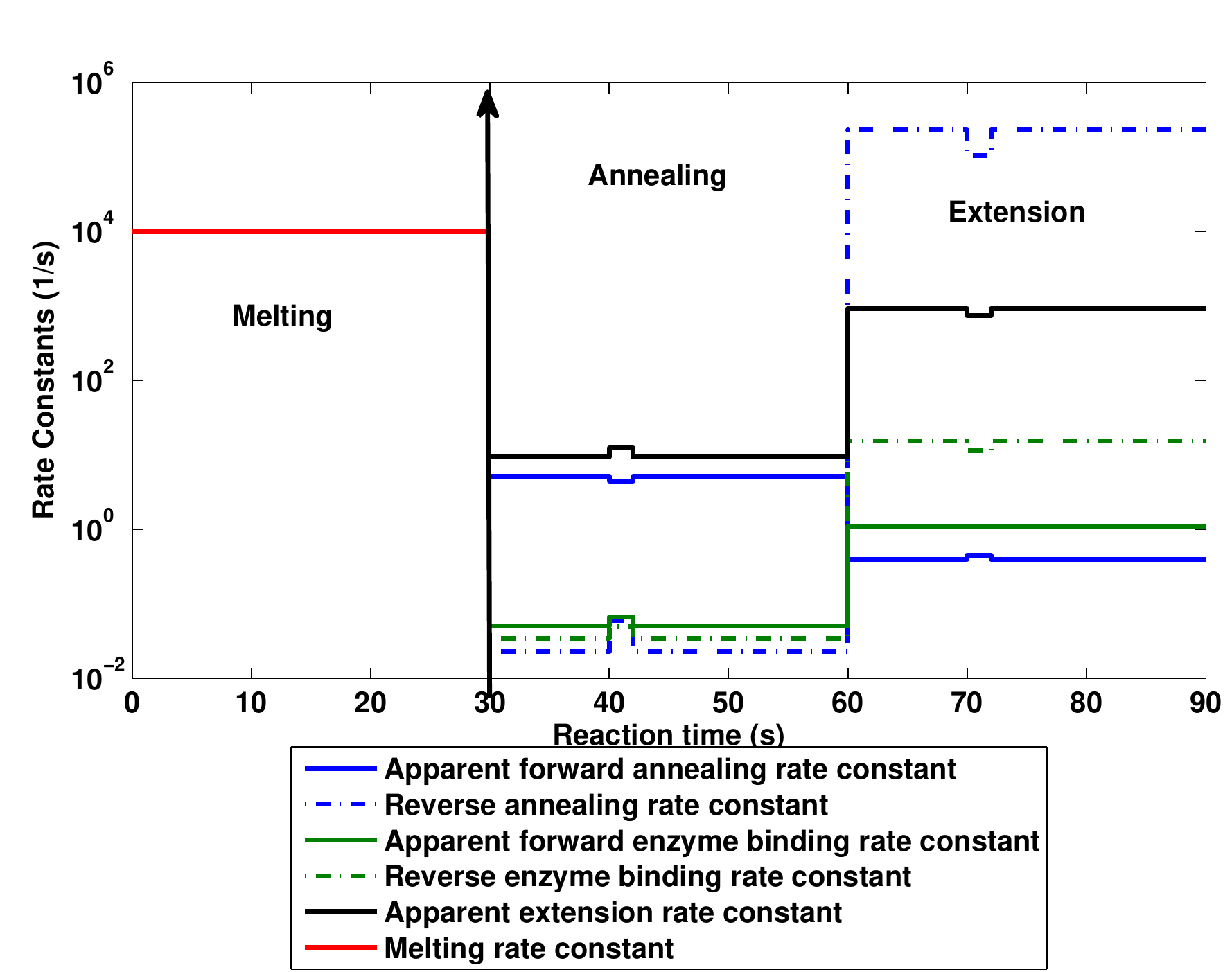}
\caption{
Temperature Variation of the DNA Amplification Rate constants. Annealing rate constants have been obtained for a primer with 15 base pairs. The second order forward annealing rate constant, forward enzyme binding rate constant and extension rate constants have been multiplied with Primer concentration (1 $\mu$M), Enzyme concentration (10 nM) and Nucleotide Concentration (100 $\mu$M). Annealing and extension temperatures are assumed to be 35 and 72 $^0$C, respectively. The step changes in rate constants depict the effects of change in 2 $^0$C in temperature. During the melting step all other rate constants have been assumed to be zero and the melting rate constant is assumed to be $10^4$.
}
\label{k_comparison}
\end{figure}

As we have discussed above, Mehra and Hu \cite{Mehra} developed a staged time invariant model. For the same model structure, one might vary the rate constants in each step to develop a staged time varying model. The staged time varying PCR model is a special case of a time varying model with  drift. In a state space model, if a specific set of control variables are kept constant, then the type of state space system is called a time-varying model with drift. During the melting step, the control variables corresponding to the other two steps can be kept constant and the corresponding approximations can be made in the annealing and extension steps as well. Table \ref{model_comparison_table} presents such a model structure. The accuracy of drift approximations can be evaluated by comparison of the relative magnitudes of the reaction rate constants at specified temperatures, which are presented in Fig. \ref{k_comparison}. It can be seen that if drift is set to zero in the above model, it reduces to a staged time varying PCR model. Note that drift vector fields for DNA amplification control systems are sequence-dependent due to sequence dependence of the associated $k_i$\rq{}s.
\subsection{Time Varying DNA Amplification Model (TVM)}
The time-varying PCR model allows the melting, annealing, and enzyme binding/extension vector fields to be applied simultaneously and allows time varying manipulation of the corresponding temperature-dependent rate constants described in Section \ref{model_develop}. $x\left(t\right)$ and $g_i\left(x\left(t\right)\right)$ are the same as in the time invariant model. The only change here is that $k\left(t\right)$ is a function of time. Since the  rate constants vary with respect to time, this model structure is useful for optimal control calculations that provide the optimal temperature profile by considering the whole state space. Hence, fully time-varying models do not require specification of annealing and extension steps in advance.

\begingroup
\squeezetable
\begin{table*}[ht]
\caption{Classification of DNA amplification control systems}
\centering
\begin{tabular}{c c c c c c c c c}
\hline\hline
  & \multicolumn{2}{c}{ STIM} & \multicolumn{2}{c}{TIM} & \multicolumn{2}{c}{ TVM} & \multicolumn{2}{c}{TVMD} \\
\hline
Reaction Steps & $g_0\left(x\right)$ & $f\left(x,k\right)$ & $g_0\left(x\right)$ & $f\left(x,k\right)$ & $g_0\left(x\right)$ & $f\left(x,k\right)$ & $g_0\left(x\right)$ & $f\left(x,k\right)$  \\ [0.5ex]
\hline
Melting & 0 & $\sum_{i = 1}^2 k_i g_i\left(x\left(t\right)\right)$ & 0 & $\sum_{i = 1}^{10} k_i g_i\left(x\left(t\right)\right)$ & 0 & $\sum_{i = 1}^{10} k_i\left(t\right) g_i\left(x\left(t\right)\right)$ & $\sum_{i = 3}^{10} k_i g_i\left(x\left(t\right)\right)$ & $\sum_{i = 1}^2 k_i g_i\left(x\left(t\right)\right)$ \\
Annealing & 0 & $\sum_{i = 3}^6 k_i g_i\left(x\left(t\right)\right)$ & 0 & $\sum_{i = 1}^{10} k_i g_i\left(x\left(t\right)\right)$ & 0 & $\sum_{i = 1}^{10} k_i\left(t\right) g_i\left(x\left(t\right)\right)$ & $\sum_{i = 1,2,7}^{10} k_i g_i\left(x\left(t\right)\right)$ & $\sum_{i = 3}^6 k_i g_i\left(x\left(t\right)\right)$ \\
Extension & 0 & $\sum_{i = 7}^{10} k_i g_i\left(x\left(t\right)\right)$ & 0 & $\sum_{i = 1}^{10} k_i g_i\left(x\left(t\right)\right)$ & 0 & $\sum_{i = 1}^{10} k_i\left(t\right) g_i\left(x\left(t\right)\right)$ & $\sum_{i = 1}^8 k_i g_i\left(x\left(t\right)\right)$ & $\sum_{i = 7}^{10} k_i g_i\left(x\left(t\right)\right)$ \\[0.5ex]
\hline\hline \\
\end{tabular}
\label{model_comparison_table}
\end{table*}
\endgroup

\section{Comparison of model-predicted DNA amplification dynamics}
\begin{figure} 
\centering
\includegraphics[width=8cm,height=6cm]{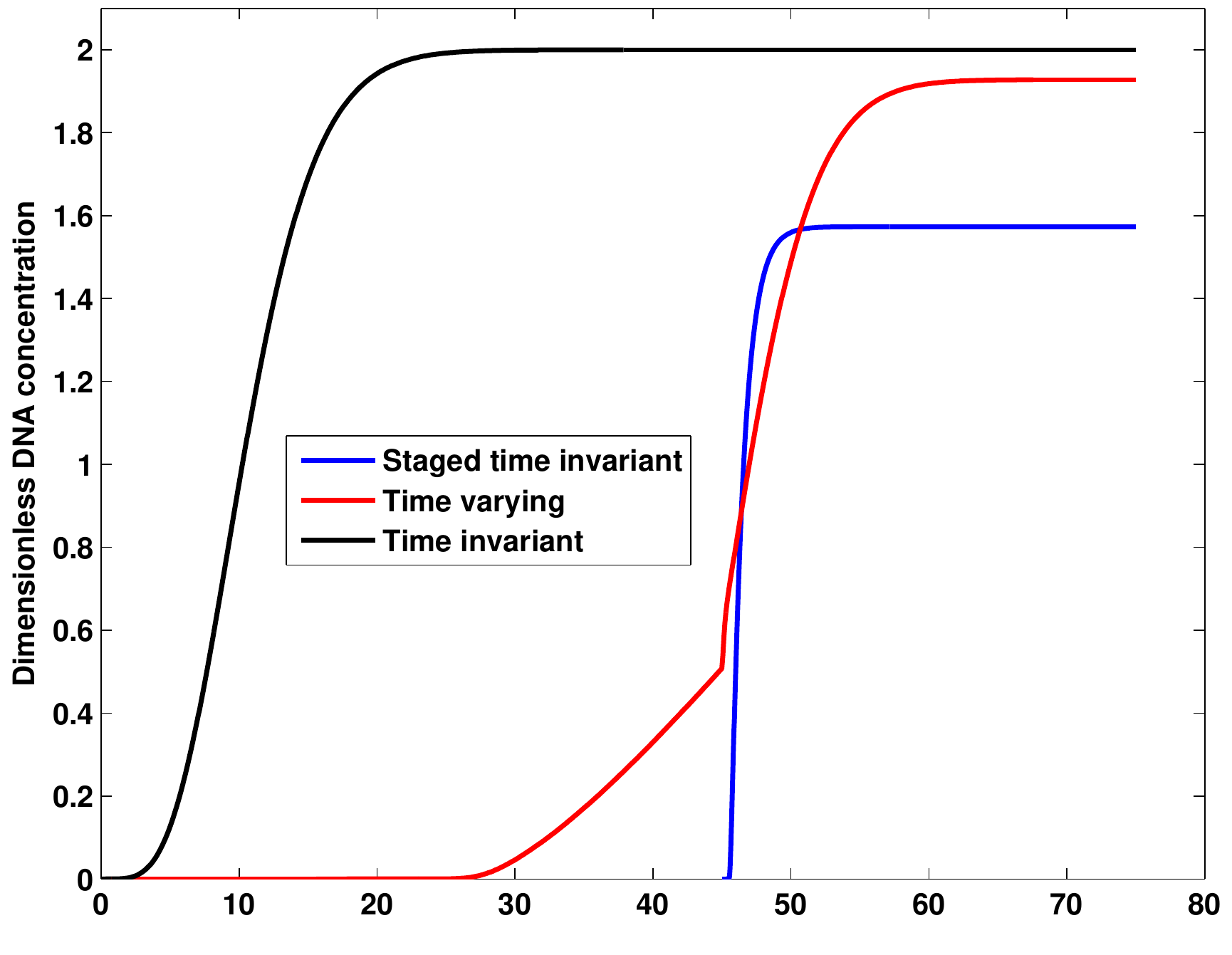}
\caption{Comparison between staged time-invariant (on-off), time-invariant and time-varying models. Annealing time and temperature are fixed to be 45 seconds and 30 $^0$C. Extension time and temperature are fixed to be 30 seconds and 72 $^0$C. Initial concentrations of template, primer, enzyme, and nucleotide are 10 fM, 0.2 $\mu M$, 10 nM, 800 $\mu M$, respectively. Since the melting step is neglected in this simulation, the reaction time is sum of the annealing and extension times only. The time-varying model with drift is not shown since the choice of drift vector field is not unique.}
\label{model_comparison}
\end{figure}

Several of the control systems introduced above (STIM, TIM, TVMD) make approximations regarding the magnitudes and temperature variation of the rate constants $k_i$ that enable the corresponding vector fields to be treated either as drift or to be turned off during certain steps of PCR. In temperature-controlled DNA amplification, where $k_i=k_i(T)$, these approximations may not be valid for arbitrary DNA sequences. The variation of melting, annealing, enzyme binding and extension rate constants with respect to time in each step of a standard PCR protocol, depicted in Fig. \ref{k_comparison}, clearly indicates simultaneous annealing, enzyme binding and extension. The effects of 2 $^0$C changes in the temperatures of each step are depicted to assess the validity of drift approximations. Fig. \ref{model_comparison} compares the evolution of the DNA concentration in a first PCR cycle based on the above models. It is clear from Fig. \ref{model_comparison}  that the time invariant model is unrealistic as it completes the annealing and extension within 20 s during the annealing step. However, the predictions of the time-varying model are consistent with the general Real-Time PCR temperature cycling prescription. The staged time invariant model does not account for the E.Di dissociation and enzyme binding during the annealing step. As a result of this, it can be seen in Fig. \ref{model_comparison} that the whole extension reaction is completed within 5 seconds and this contradicts the general Real-Time PCR experimental observations.

Therefore, we conclude that the time-varying PCR model is required for dynamic optimization of reaction conditions that can exploit the simultaneous reactions that have been demonstrated to play an important role in PCR dynamics \cite{biophys2014}. Proper modeling of temperature-controlled DNA amplification requires equality constraints to be applied to the components of the vector of rate constants $k$, based on biophysical modeling of the temperature dependence of the rate constants. We impose these constraints in Section \ref{fixed_opt_prob}.

\section{Suboptimally Controlled Geometric Amplification of DNA} \label{simplex}
Standard PCR cycling strategies are based on approximations (like the STIM in Table \ref{model_comparison_table}) to the true dynamics of PCR. In this section, we show that these cycling strategies are suboptimal given the actual sequence- and temperature-dependent dynamics. We simulate the geometric growth of DNA concentration, using the sequence- and temperature-dependent TVM model for DNA amplification, for various choices of manipulated inputs, establishing lower bounds on the margin of improvement that can be achieved through use of optimal cycling strategies.

Figs \ref{temp_time_profile_45s_30s} and \ref{temp_time_profile_120s_30s} depict the temperature cycling protocols considered.
In \cite{biophys2014}, the importance of simultaneous annealing and extension (vector fields $g_3$ to $g_6$ and $g_7$ to $g_{10}$, respectively) was demonstrated through transient dynamics simulations of single PCR cycles.
Generally, lower annealing temperatures increase the primer hybridization efficiency, but higher annealing temperatures increase the polymerase binding and extension rates. The initial species concentrations and reaction time determine which effect dominates. Here, we examine the net effect of these vector fields on the geometric growth rate of DNA concentration using the aforementioned choices of manipulated inputs, illustrating the role played by the initial species concentrations at the start of a cycle (which vary with the stage of PCR) in determining the growth rate. The following primers were used in the study: P1: GCTAGCTGTAACTG ($T_m=40^0$C) and P2: GTCTGCTGAAACTG ($T_m=42^0$C). 

\subsection{Geometric growth of DNA at low reaction time} \label{cyclic_profile_stage1}
The geometric growth of DNA concentration in Fig. \ref{gg_growth} is similar to that of a typical real time PCR. In Fig. \ref{gg_growth}, the aim is to maximize the concentration of DNA at a specified time by modifying the temperatures of the PCR reaction steps.
At 50 $^0$C, as shown in Fig. \ref{gg_growth}, the efficiency is much lower than that at 30 to 45 $^0$C, due to the lower melting temperature of the primers.

When the enzyme concentration is in large excess compared to the DNA concentration during the initial stage of PCR, the enzyme binding reaction is a pseudo-first order reaction. Therefore, the evolution of the DNA concentration in a particular cycle in this stage does not depend on enzyme concentration.
Due to the comparable enzyme and target DNA concentrations in the second stage of PCR, which leads to a second-order enzyme binding reaction after 24 cycles, the lower enzyme binding and extension rates reduce the efficiency at 30 $^0$C annealing temperature.
At 45 $^0$C, the primer hybridization efficiency is lower; this effect dominates and decreases the geometric growth rate in the initial stage of PCR, but the higher polymerase binding and extension rates dominate in the later cycles of PCR, resulting in a final DNA concentration similar to that at 40 $^0$C after 29 cycles.  Once the DNA concentration is equal to the enzyme concentration in the second stage of PCR, the latter will become the limiting reactant.
Given more reaction time, the enzyme could bind to and polymerize excess $SP$ templates after it dissociates from fully-extended dsDNA.

\subsection{Geometric growth of DNA at high reaction time} \label{cyclic_profile_stage2}
\begin{figure*} 
\centering
\subfigure[]{
\includegraphics[width=7cm,height=5cm]{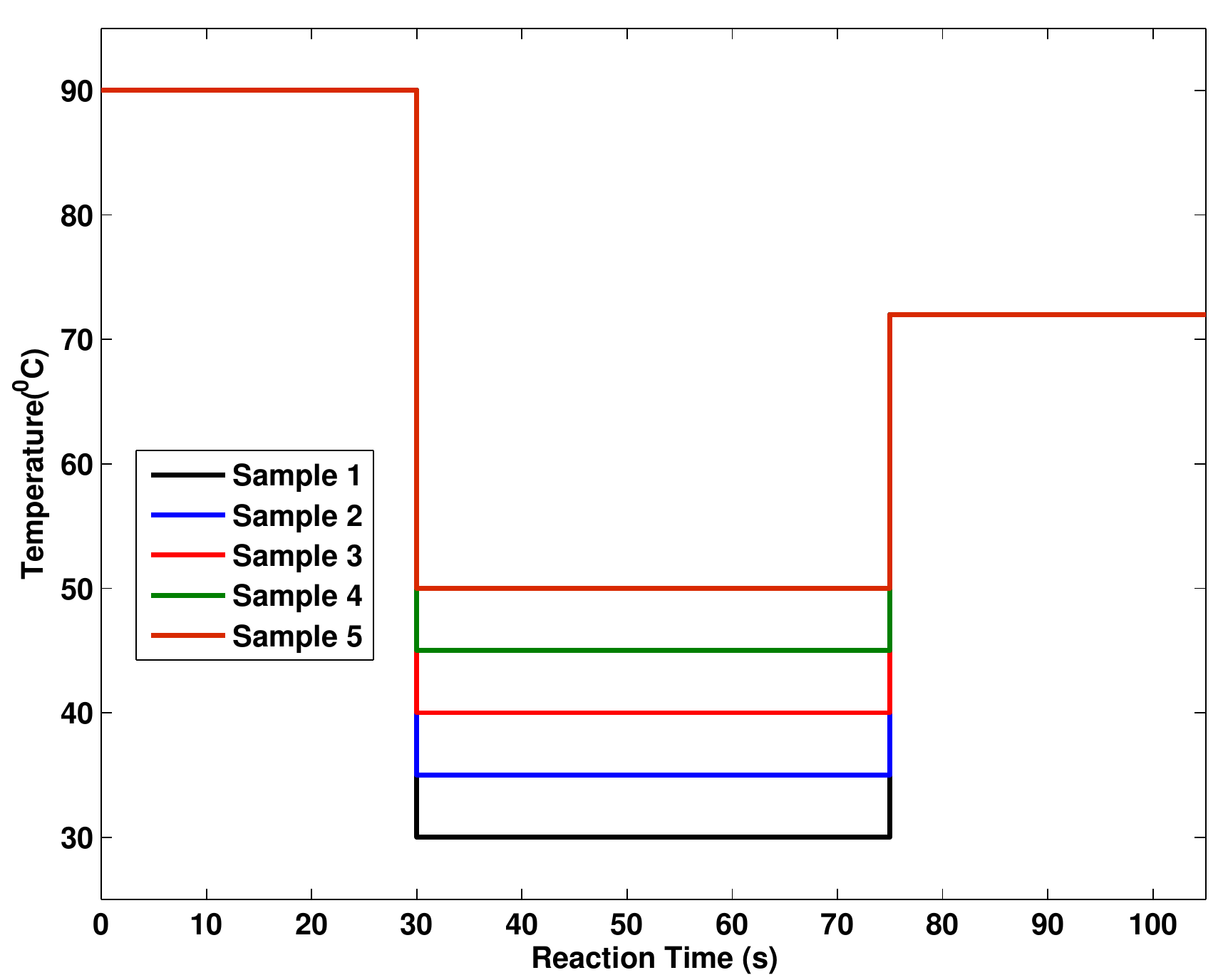}
\label{temp_time_profile_45s_30s}}
\quad
\subfigure[]{
\includegraphics[width=7cm,height=5cm]{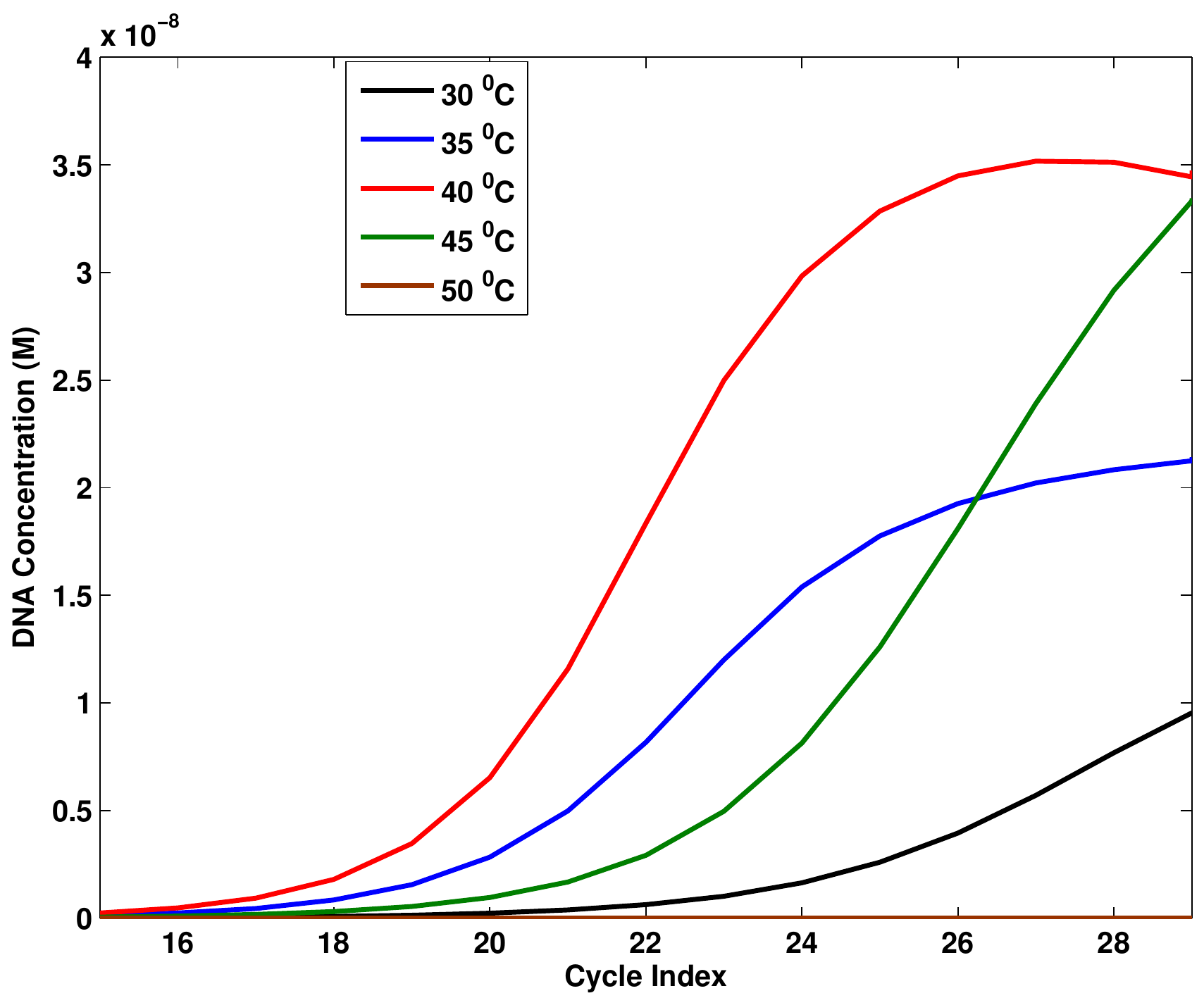}
\label{gg_growth}}
\quad
\subfigure[]{
\includegraphics[width=7cm,height=5cm]{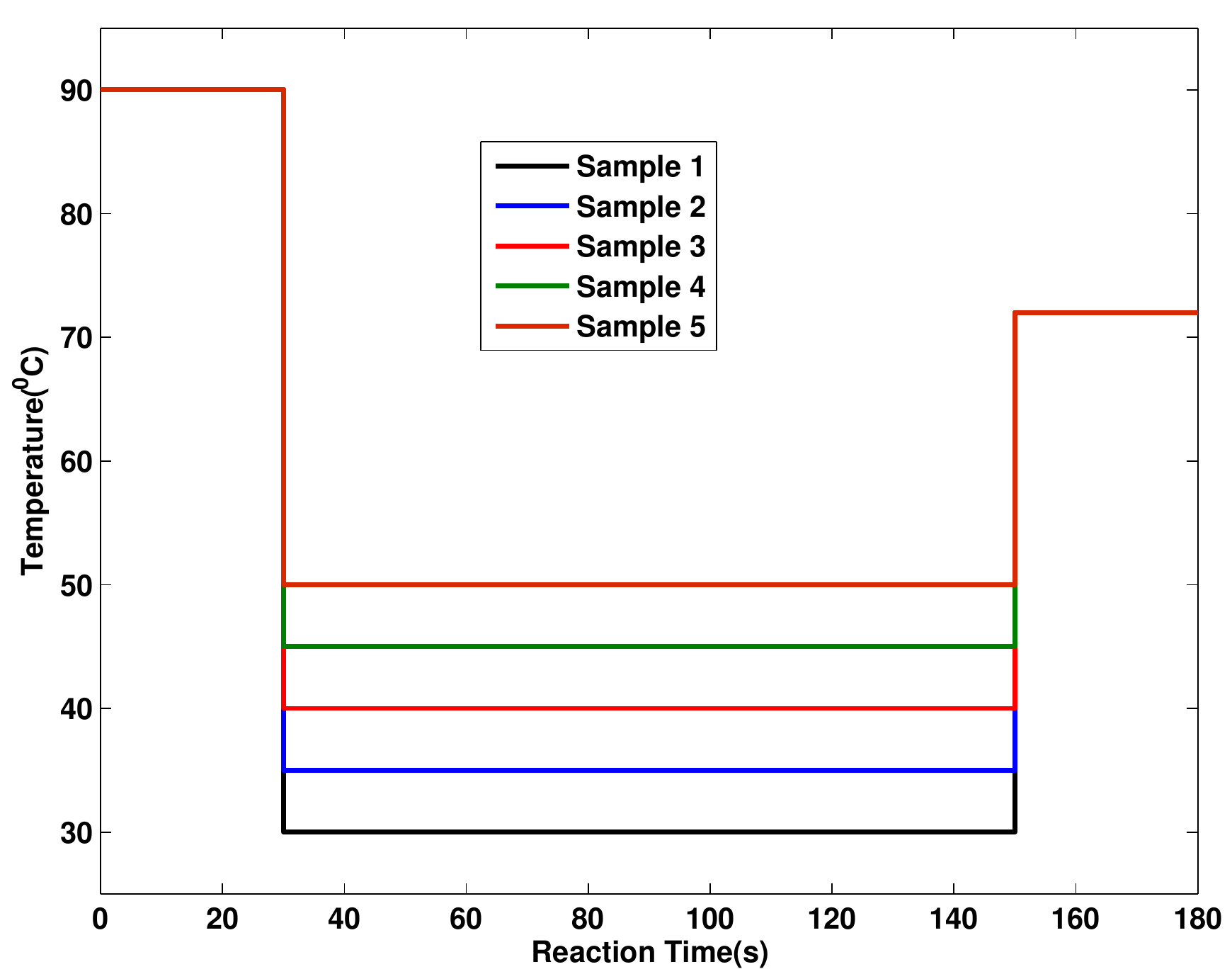}
\label{temp_time_profile_120s_30s}}
\quad
\subfigure[]{
\includegraphics[width=7cm,height=5cm]{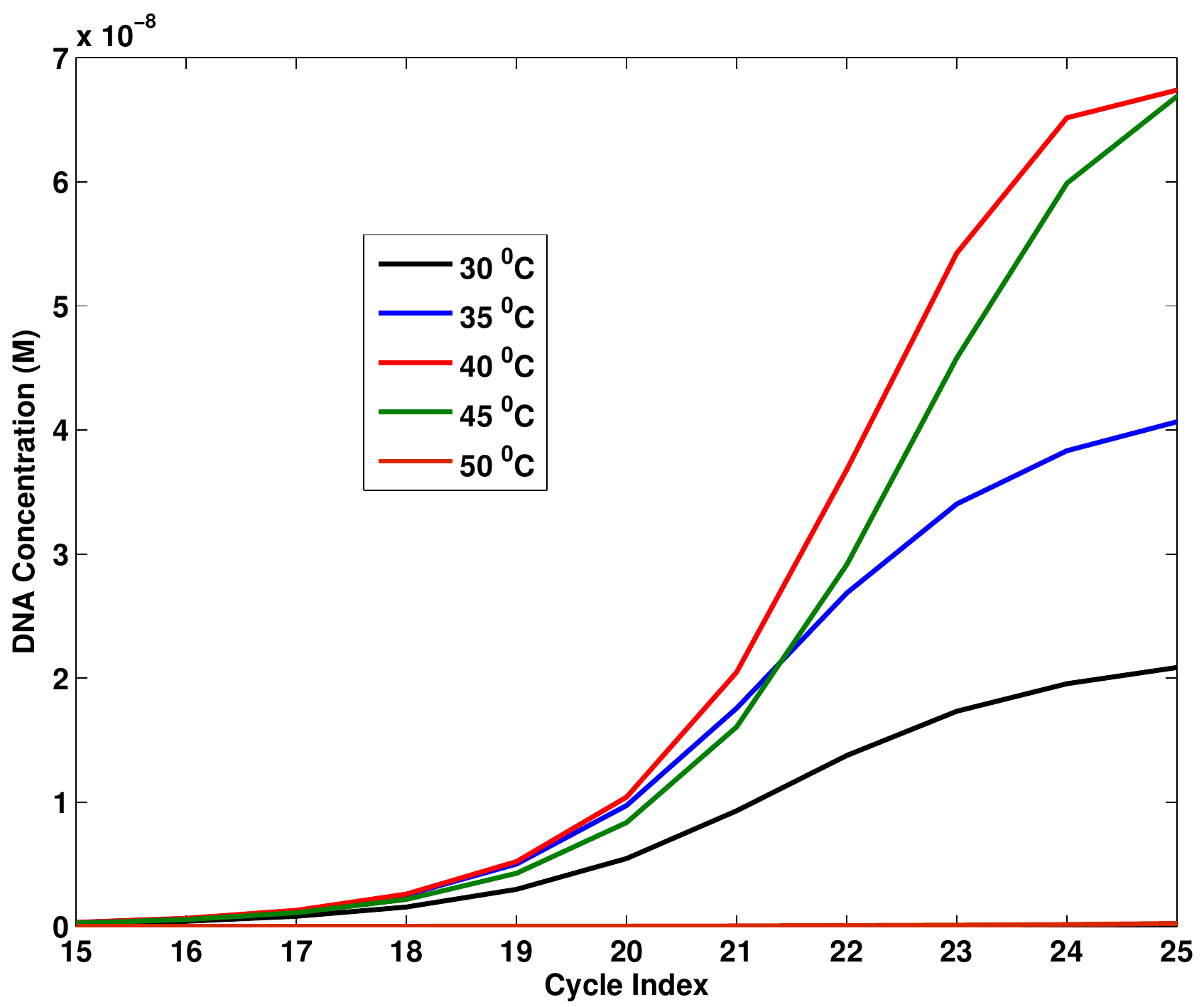}
\label{gg_growth_prolonged}}
\caption{Initial concentrations of template, primer, enzyme, and nucleotide are 10 fM, 0.2 $\mu M$, 10 nM, 800 $\mu M$, respectively and the annealing temperature is varied from 30 $^0$C to 50 $^0$C using grid based sampling. a) Temperature vs Time profile for an annealing time of 45 seconds and extension time of 30 seconds. b) Geometric growth of DNA with shorter reaction reaction time (45 s of annealing time). c) Temperature vs Time profile for an annealing time of 120 seconds and extension time of 30 seconds. d) Geometric growth of DNA with prolonged reaction reaction time (120 s of annealing time).
}
\label{grid_sampling}
\end{figure*}

The maximum amount of DNA that can be obtained from PCR is equal to the initial concentration of primers. As shown above, in the final stage of PCR, when the overall reaction time is low the maximum DNA concentration that is obtained is less than the initial concentration of primers. This suggests that there can be some improvement in the final DNA concentration if the reaction times are also changed appropriately. Therefore, the annealing time was increased to 120 seconds. As presented in Fig. \ref{gg_growth_prolonged}, with this change, the DNA concentration after 25 cycles is higher than that of the previous study. A maximum concentration of 70 nM at 40 $^0$C is obtained at the end of 25 cycles when annealing time is increased to 120 seconds. 
When the reaction time is increased at higher annealing temperature (45 $^0$C), more DNA is produced after 25 cycles than at lower temperature.

Once the enzyme has become the limiting reactant, unless the enzyme molecules are released after converting the equivalent number of ssDNA into dsDNA, all the ssDNA cannot be converted into  dsDNA. As noted above, since the enzyme binding and extension reactions are faster at higher annealing temperatures, these temperatures produce more DNA than lower annealing temperatures. 

Thus, from this study, we conclude that a systematic procedure should be formulated to obtain the optimal annealing time and temperature. Moreover, due to geometric growth, there can be a considerable difference between the final DNA concentrations produced by control strategies ($T(t)$) based on misspecified and properly specified DNA amplification models.

\section{Control Strategy} \label{control_strategy}

In the previous section, we introduced control systems for DNA amplification and studied the dynamics of geometric growth of DNA for several types of manipulated inputs. We showed that when the annealing time and temperature were varied, due to geometric growth, there was a considerable difference in the final DNA concentration. 
Therefore, the determination of optimal time and temperature is essential. In this section we show how to formulate and solve an optimal control problem to obtain not just the optimal annealing time and temperature, but the optimal PCR temperature cycling strategy for a particular amplification objective.

\subsection{Optimal Control Problem formulation} \label{optimal_control_formulation}
In  optimal control, the optimal evolution of a manipulated variable is sought by minimizing or maximizing a desired objective function. In a PCR, the manipulated variable is the reaction temperature and the desired objective is to maximize the target DNA concentration at the end of every cycle. We classify PCR optimal control problems into two types as follows:

\begin{itemize}
\item \textit{Fixed time optimal control problem} -  Here we obtain the optimal temperature profile that maximizes the desired DNA concentration profile in a given fixed reaction time. If the cycle time is known in advance, instead of following a grid based optimization approach as shown in Fig. \ref{gg_growth}, it is possible to obtain the optimal temperature profile by solving a fixed time optimal control problem.
\item \textit{Minimal time optimal control problem} - Here we obtain the optimal temperature profile that minimizes the overall reaction time for achievement of a specified level of amplification. This allows automated determination of the cycle time and avoids the need for sampling cycle times as was shown in Figs. \ref{temp_time_profile_45s_30s} and \ref{temp_time_profile_120s_30s}.
\end{itemize}

The desired objective can be expressed in the general form:
\begin{align}
J = F(x(t_f)) + \int_0^{t_f} L(x(t),u(t)) dt
\end{align}
$F(x(t_f))$ is referred to as an endpoint or Mayer cost and $\int_0^{t_f} L(x(t),u(t)) dt$ is referred to as a Lagrange cost.

\subsection{Fixed-time DNA amplification optimal control problem} \label{fixed_opt_prob}
In Fig. \ref{gg_growth}, for a fixed reaction time, we varied the reaction temperature to maximize the DNA concentration. This grid-based sampling approach is not an efficient way to obtain the optimal temperature profile. Therefore, we solve an optimal control problem with a desired objective function that maximizes the DNA concentration. Typically a lower limit ($T_{min}$) for the annealing temperature is applied to reduce the operating temperature range. The upper limit of the operating temperature range could be the maximum melting temperature ($T_{max}$). Hence lower and upper limits for the PCR reaction temperature can be applied and this can be expressed as an inequality constraint for the optimal control problem. Thus the following optimal control problem can be formulated to maximize the target DNA concentration at the end of each cycle:
\begin{align}
\underset{{T(t)}}{\text{min}} \hspace{2pt} J &= \sum_{i =1}^{4n + 4} x_i \left(t_f\right) \label{obj_fun_ini} \\
s.t. ~~\hspace{2pt} \frac{dx(t)}{dt} &=\sum_{i=1}^{10} k_{i,seq}\left(T\left(t\right)\right) g_i\left(x\left(t\right)\right), \hspace{2mm} ~~x(0) = x_0  \label{state_cons} \\
 T_{min}&\leq T(t) \leq T_{max} \label{T_limit}\\
 x,~g_i\left(x\right) &\in \mathbb{R}^{4n+9},~x \geq 0,\\
 h\left(x\right) &\in  \mathbb{R}^{5}, ~h\left(x\right) =0
\end{align}
\noindent
where $k_{i,seq}(T)$ denote the sequence- and temperature-dependent rate constants described in Section \ref{model_develop}.
$i$ in the objective function (Eq. (\ref{obj_fun_ini})) indexes all the state variables except for $DNA$, $P_1$, $P_2$, $E$, and $N$.
Note that due to the equality constraints $h(x)=0$ on the state vector, $\sum_{i=1}^{4n + 4} x_i \left(t_f\right)$ uniquely determines $x_{4n+5}$, which is the DNA concentration, such that minimization of the former maximizes the latter. Eq. (\ref{obj_fun_ini}) specifies a Mayer functional with $F(x(t_f)) = \sum_{i=1}^{4n+4} x_i(t_f)$ and $L(x(t),u(t))=0$.

The solution of the above optimal control problem provides the optimal trajectory for temperature ($T^*\left(t\right)$) and concentration ($x^*\left(t\right)$) profiles with respect to time.  $t_f$ in Eq. (\ref{obj_fun_ini}) can correspond to one or multiple cycles; in the former case, assumes a fixed total reaction time for that cycle is specified in advance.
Note that in this formulation, the control (manipulated input) variable is the temperature $T$, since in PCR reactions the rate constants are varied through temperature cycling. This variable appears in the state equations through the rate constants. Eq. (\ref{obj_fun_ini}) assumes a fixed total reaction time for a given cycle. Since the objective function is a Mayer functional, there are multiple solutions possible for the above optimal control problem.

\subsubsection{Rate constant as a control variable: Relationship between PCR Rate Constants}
The relationship between the temperature and the rate constants is typically given by the Arrhenius equation 
\begin{align}
k_i = k_{0i} \exp\left(\frac{-E_{ai}}{RT}\right)  \label{arrhenius}
\end{align}
$E_a$ is the activation energy and $k_0$ is the pre-exponential factor. $E_a$ represents the apparent activation enthalpy of the reaction, whereas $k_0$ is a function of the apparent activation entropy of the reaction. In general, these thermodynamic parameters may vary with temperature. Temperature-dependent apparent activation enthalpy and entropy can be accommodated through generalizations of Eq.  (\ref{arrhenius}). However, we have shown in \cite{biophys2014} that Eq. (\ref{arrhenius}) can provide a reasonable approximation for the temperature variation of PCR reaction rate constants.

Though in Eq. (\ref{state_cons}) we specified temperature as a control variable, as we see in Eq. (\ref{arrhenius}) it appears in the exponential term. This may introduce a strong nonlinearity and hence causes computational difficulty to solve this optimal control problem. In order to avoid this issue, it is possible to use all these 5 rate constants as the control variables and find their optimal evolution. In this formulation, the optimal control problem can be re-written as
\begin{align}
\underset{{k(t)}}{\text{min}} \hspace{2pt} J &= \sum_{i=1}^{4n+4} x_i\left(t_f\right) \label{u_manipulated_1} \\
\frac{dx(t)}{dt} &=\sum_{i=1}^{10} k_i(t) g_i(x(t)), \hspace{2mm}~~ x(0) = x_0 \label{u_manipulated_2}  \\
x,~g_i\left(x\right) &\in \mathbb{R}^{4n+9}, ~x \geq 0 \\
k &\in \mathbb{R}^{10},~k \geq 0,~k_{1,seq}^{min}  \leq k_{1}\left(t\right) \leq k_{1,seq}^{max}   \\
k_{j}  &= \alpha_{j,seq} k_{1}^{\beta_{j,seq}}, \hspace{5mm} j = 2,...,10  \\
\alpha_{j,seq} &= \left(k_{0j,seq}/k_{01,seq}\right)^{\beta_{j,seq}},~\beta_{j,seq} = E_{aj,seq}/E_{a1,seq} \nonumber\\
h\left(x\right) &\in  \mathbb{R}^{5},~h\left(x\right) =0
\end{align}
\noindent
Where $k_{1,seq}^{min} $ and $k_{1,seq}^{max}$ are the lower and upper bounds for the rate constants $k_{1,seq}$ correspond to $T_{min}$ and $T_{max}$. The subscript $seq$ indicates the sequence dependency. The control system defined by Eq. (\ref{u_manipulated_2}) is a special case of Eq. (\ref{seq_gform}). Thus, the optimal control and concentration trajectory is denoted as $\left(k^*_{seq} \left(t\right),  x^*_{seq} \left(t\right) \right)$. This trajectory can then be mapped onto the optimal trajectory $(T_{seq}^*(t), x_{seq}^*(t))$ using Eq. (\ref{arrhenius}).

\subsection{DNA amplification in minimal time: Time Optimal Control} \label{time_opt_control}
In Fig. \ref{gg_growth} and Fig. \ref{gg_growth_prolonged} an arbitrary reaction time has been chosen to obtain the DNA concentration profile. Even though the amplification efficiency at certain temperatures is nearly 100 \%, the cycle reaction time and hence the overall reaction time (sum of reaction time of all the PCR cycles) can be further reduced. To achieve a specified level of DNA amplification in the shortest possible time, the optimal control problem should be formulated in such a way that the solution minimizes the overall reaction time. 

Minimization of the overall reaction time for DNA amplification can be achieved through a minimum time optimal control framework. In particular, the time optimal control framework provides prescriptions for the optimal cycle switching time on a formal mathematical basis.

In time optimal control, in addition to state variables, evolution time also must be optimized. A Lagrange cost of the form $\int_0^{t_f} dt$  corresponding to the evolution time is used as an objective function and state vector is constrained to achieve a specified level of DNA amplification at the end of a cycle as follows:
\begin{align}
\underset{{k(t)}}{\text{min}} \hspace{2pt} J &= \int_0^{t_f} dt \label{obj_fun_2} \\
s.t. \frac{dx(t)}{dt} &=\sum_{i=1}^{10} k_i(t) g_i(x(t)), \hspace{2mm} ~~x(0) = x_0\\
\hspace{2mm}  x_{4n+5}\left(t_f\right) &= x_{4n+5,f} \geq 2^mx_{4n+5}(0),~~m \in \mathbb{Z}^{0+} \label{end_point_cons} \\
 x,~g_i\left(x\right) &\in \mathbb{R}^{4n+9},~x \geq 0,\\
k &\in \mathbb{R}^{10},~k \geq 0, ~k_{1,seq}^{min}  \leq k_{1}\left(t\right) \leq k_{1,seq}^{max}\\
k_{j}  &= \alpha_{j,seq} k_{1}^{\beta_{j,seq}}, \hspace{5mm} j = 2,...,10  \\
\alpha_{j,seq} &= \left(k_{0j,seq}/k_{01,seq}\right)^{\beta_{j,seq}} \nonumber\\
~\beta_{j,seq} &= E_{aj,seq}/E_{a1,seq} \nonumber\\
 h\left(x\right) &\in  \mathbb{R}^{5}, ~h\left(x\right) =0
\end{align}
Note that $x_{4n+5} = x_1^0 + x_{2n+3}^0 - \sum_{i=1}^{4n+4} x_i$. Where $x_1^0$ and $x_{2n+3}^0$ are the initial concentrations of first and second single strand. $m$ denotes the number of cycles for which the time optimal solution is required. Thus, if $m=1$, $x_{4n+5,f} \geq 2 x_{4n+5}(0)$, which implies that the time is minimized for amplification efficiency greater than $1$, which in turn implies that more than one cycle will be generated. The single cycle time optimal solution is obtained by considering $k(t)$ up to the switching time between cycles.
In addition to the regular first order conditions for optimality, additional optimality conditions must be satisfied for the time optimal control problem (Eq. (\ref{obj_fun_2})) and these are described in \cite{Stengel}.

\subsection{A Strategy for Optimal Synthesis of the DNA Amplification Control Trajectory: Stage 1}


From the enzyme concentration and the evolution of DNA concentration throughout the amplification reaction, the reaction can be separated into two stages. Stage 1 corresponds to a resource-unlimited environment for sequence replication.  Stage 2 begins when environmental resource limitations affect the probability of sequence replication.

Stage 1 is comprised of all the cycles at which the enzyme concentration is higher than the target DNA concentration by 2 orders of magnitude so that a pseudo first order kinetics for the enzyme binding reaction can be assumed.
Using grid-based sampling of experimental conditions, we have shown that the annealing time for this stage could be around 45 s at a specific annealing temperature. In order to find the optimal time and temperature protocol that improves upon the results in Fig. \ref{gg_growth} (cycles 1 to 15) in a systematic manner, a fixed time optimal control problem can be formulated (Eq. (\ref{u_manipulated_1})). Alternatively, it is possible to formulate the time optimal control problem (Eq. (\ref{obj_fun_2})) wherein the objective is the minimization of the reaction time for a specified target DNA concentration.  For both problems, an important feature of the solution $k^* \left(t\right)$ for multiple cycles is that they are periodic under the pseudo-first order approximation.
This reduces the computational complexity of the optimal control synthesis for Stage 1.
For example, for control problems (\ref{obj_fun_ini}) and (\ref{obj_fun_2}) with the cycle time or efficiency specified in the problem formulation, the optimal temperature profile for cycles in this Stage need only be computed once. Similarly, for control problems with evolution times or target efficiencies spanning multiple cycles in Stage 1, solutions to smaller subproblems can be used in the construction of the control trajectories.

Here we consider the control strategy for a Stage 1 minimal time control problem where the temperatures as well as switching times for the denaturation and annealing steps are constrained such that $k(t)=f(t)$. Note that application of this constraint results in a control system of the STIM type, wherein the only manipulated input parameter is the switching time between the cycles. However, the vector fields applied in each step are $\sum_{i=1}^{10} k_i(T_{1})g_i(x),~ \sum_{i=1}^{10} k_i(T_{2})g_i(x)~\sum_{i=1}^{10}, k_i(T_{3})g_i(x)$, respectively, where $T_1,T_2,T_3$ denote the temperatures of the melting, annealing and extension steps, rather than the vector fields listed for the STIM in Table \ref{model_comparison_table}.

\subsubsection{Optimal cycle switching time for control of geometric growth} \label{min_ext_time}
In this section, for a given temperature profile and hence rate constant trajectory $k(t)$ we find the optimal switching time between cycles. For example, in Fig \ref{gg_growth} for a fixed reaction time of 105 seconds, at three different temperatures we have estimated the target DNA concentration profile. Now we will analyze those DNA concentration profiles and estimate the optimal cycle time that minimizes the overall reaction time in stage 1 of the PCR. Since the optimal control profile is periodic in stage 1, the optimal cycle time for all the cycles in this stage will be the same.

Let $0 \leq \eta \leq 1$ be the efficiency in the first cycle and $n\left(\eta\right)$ be the number of cycles in stage 1, then

\begin{equation}
\left(1 + \eta\right)^{n\left(\eta\right)} = y \label{min_time_opt1}
\end{equation}
\noindent
where $$ y = \frac{[DNA]_n}{[DNA]_0} $$
Here, $[DNA]_n$ denotes the concentration of DNA after $n$ cycles and $[DNA]_0$ denotes the initial concentration of DNA.
From Eq. (\ref{min_time_opt1}), the number of cycles $n\left(\eta\right)$ can be expressed as follows:
\begin{equation}
n\left(\eta\right) =\frac{log\left(y\right)}{log\left(1 + \eta\right)} \label{min_time_opt2}
\end{equation}
Let $ t\left(\eta\right)$ be the time required for a cycle to achieve an efficiency of $\eta$ and the overall reaction time for stage 1 be $t_{total}(\eta)$; then
\begin{equation}
t_{total}(\eta) = n\left(\eta\right) t\left(\eta\right)= t\left(\eta\right) \frac{log\left(y\right)}{log\left(1 + \eta\right)} \label{min_time_opt3}
\end{equation}
Now we consider the following optimization problem to minimize the overall reaction time $t_{total}$:
\begin{equation}
\underset{{\eta}}{\text{min}} \hspace{2mm} t_{total}(\eta) = \underset{{\eta}}{\text{min}} \hspace{2mm} t\left(\eta\right) \frac{log\left(y\right)}{log\left(1 + \eta\right)} \label{min_time_opt4}
\end{equation}
Hence we seek $\eta$ such that
\begin{equation}
\frac{d\left( t\left(\eta\right) \frac{log\left(y\right)}{log\left(1 + \eta\right)}\right)}{d\eta} = 0 \hspace{2mm} \implies \frac{1}{t\left(\eta\right)} \frac{dt}{d\eta} = \frac{1}{\left(1+\eta\right)\left(log\left(1 + \eta\right)\right)}  \label{min_time_opt6}
\end{equation}
We solve the above equation graphically by plotting the left hand side and the right hand side with respect to $\eta$. From the given DNA concentration profile, it is possible to estimate the optimal efficiency $\left(\eta_{\min}\right)$ and hence, $t\left(\eta_{\min}\right)$ that minimizes the overall reaction time for geometric growth in stage 1, subject to the specified constraints.

Note that the computation of the optimal cycle switching time with a given STIM model, as above, does not provide the minimal cycle time that can be achieved with a TVM model. Let $t(\eta,T^*)$, where $T^*$ is time optimal input obtained from solution of problem (Eq. (\ref{obj_fun_2})), denote the optimal single cycle time for stage 1 as a function of specified cycle efficiency $\eta$.  Hence $t\left(\eta\right) \geq t\left(\eta,T^*\right)$.
Since $t\left(\eta\right) \geq t\left(\eta, T^*\right) $,
$\eta_{min}=\mathrm{arg~min}~t(\eta)$ computed above for the suboptimal $t\left(\eta\right)$ is smaller than $\eta_{min}$ for $t\left(\eta,T^*\right)$; i.e., the cycle should be run at least as long as computed based on the suboptimal $t\left(\eta\right)$ using the above model, and
\begin{equation}
\underset{{\eta}}{\text{min}} \hspace{2mm}  t_{total}(\eta) \geq \underset{{\eta}}{\text{min}} \hspace{2mm} t\left(\eta,T^*\right) \frac{log\left(y\right)}{log\left(1 + \eta\right)} \label{min_time_opt7}
\end{equation}

\begin{figure}
\includegraphics[width=8cm,height=6cm]{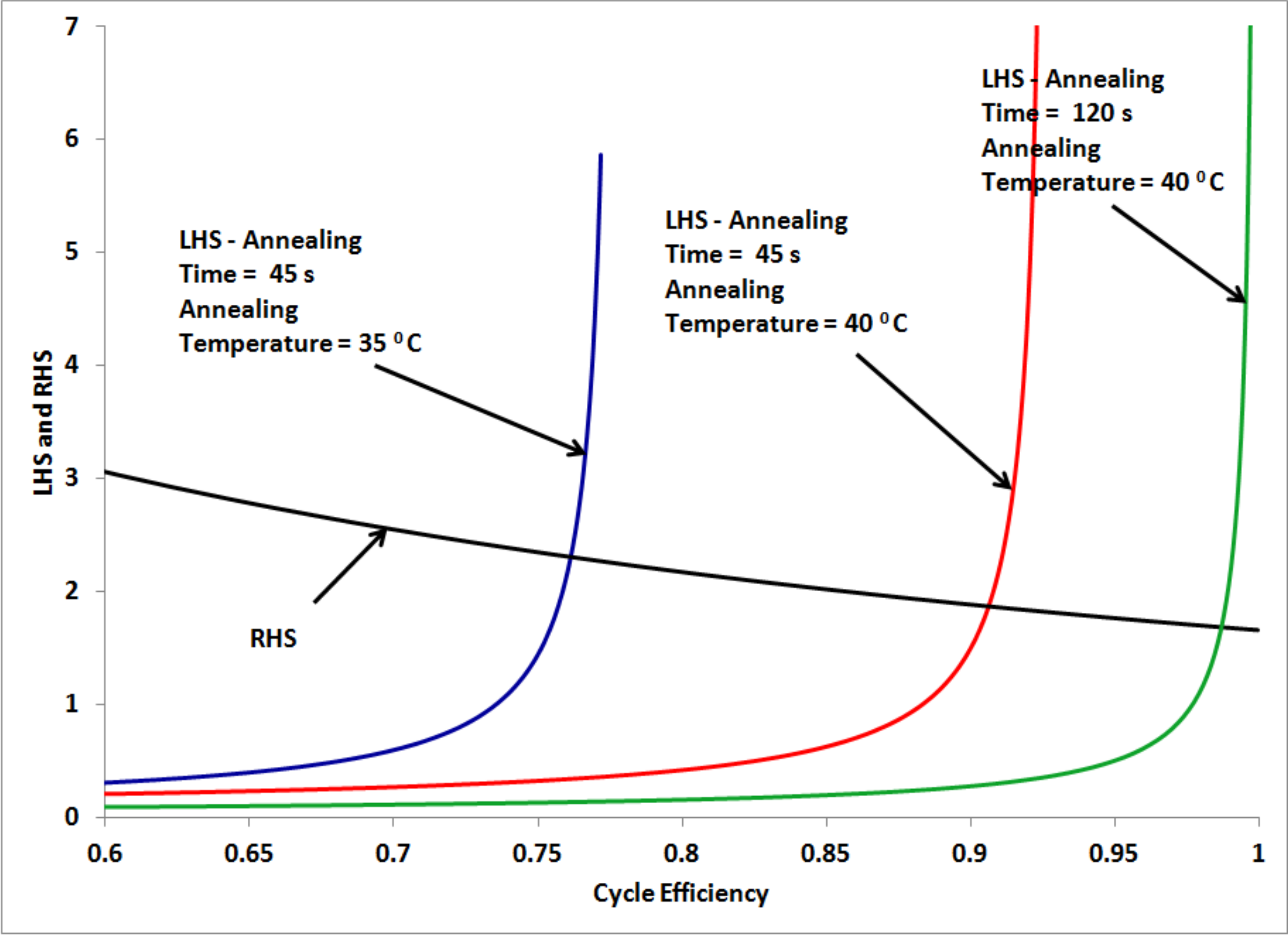}
\caption{Optimal cycle efficiency at three different reaction conditions. Temperature and time in the legend represent the annealing temperature and time. Y axis represents the LHS and RHS of Eq. (\ref{min_time_opt6}).
The intersection points specify the optimal switching times between cycles.}
\label{optimal_eta}
\end{figure}

\paragraph{Minimal reaction time example}
We consider a PCR reaction with primers and reaction conditions that are considered in Fig \ref{gg_growth} and Fig \ref{gg_growth_prolonged}. We consider annealing temperatures, 35 and 40 $^0$C from Fig. \ref{gg_growth} and 40 $^0$C from Fig. \ref{gg_growth_prolonged}. It should be noted that cycle time including the melting step in Fig. \ref{gg_growth} is 105 seconds and in Fig. \ref{gg_growth_prolonged} it is 270 seconds. Fig. \ref{optimal_eta} shows the optimal efficiency for these three conditions and from this, in each case the overall reaction time to reach the specific DNA concentration (100 nM) has been calculated. This is a constrained optimization version of the time optimal problem (Eq. (\ref{obj_fun_2})) with the additional constraint that $k(t)=f(t)$. The problem is hence only to find $t_f$ (since $k(t)$ does not change).

Table \ref{opt_eta_comparison} compares the overall reaction time under these three conditions and it can be observed that given this sample set, it is possible to reduce the overall reaction time at the specific reaction temperatures.
\begingroup
\squeezetable
\begin{table}[ht]
\caption{Comparison between optimal cyclic efficiency under different reaction conditions}
\begin{tabular}{c c c c c c}
\hline\hline
Annealing  & Annealing  & Optimal & Optimal  & Number of  & Overall \\
Temperature    &  reaction  & Efficiency& reaction  & of &reaction  (s) \\
 ($^0$C) &   time (s) &  (\%)&  time & cycles & time (s) \\  [0.5ex]
\hline
35 & 45 & 76.09 & 102.2 & 16.2 &1664 \\
40 & 45 & 90.62 &99.5 & 14.2 &1421\\
40 & 120 & 98.68 &165.3& 13.4 &2218\\ [1ex]
\hline\hline \\
\end{tabular}
\label{opt_eta_comparison}
\end{table}
\endgroup

\begin{figure*}
\centering
\subfigure[]{
\centering
\includegraphics[width=8cm,height=6cm]{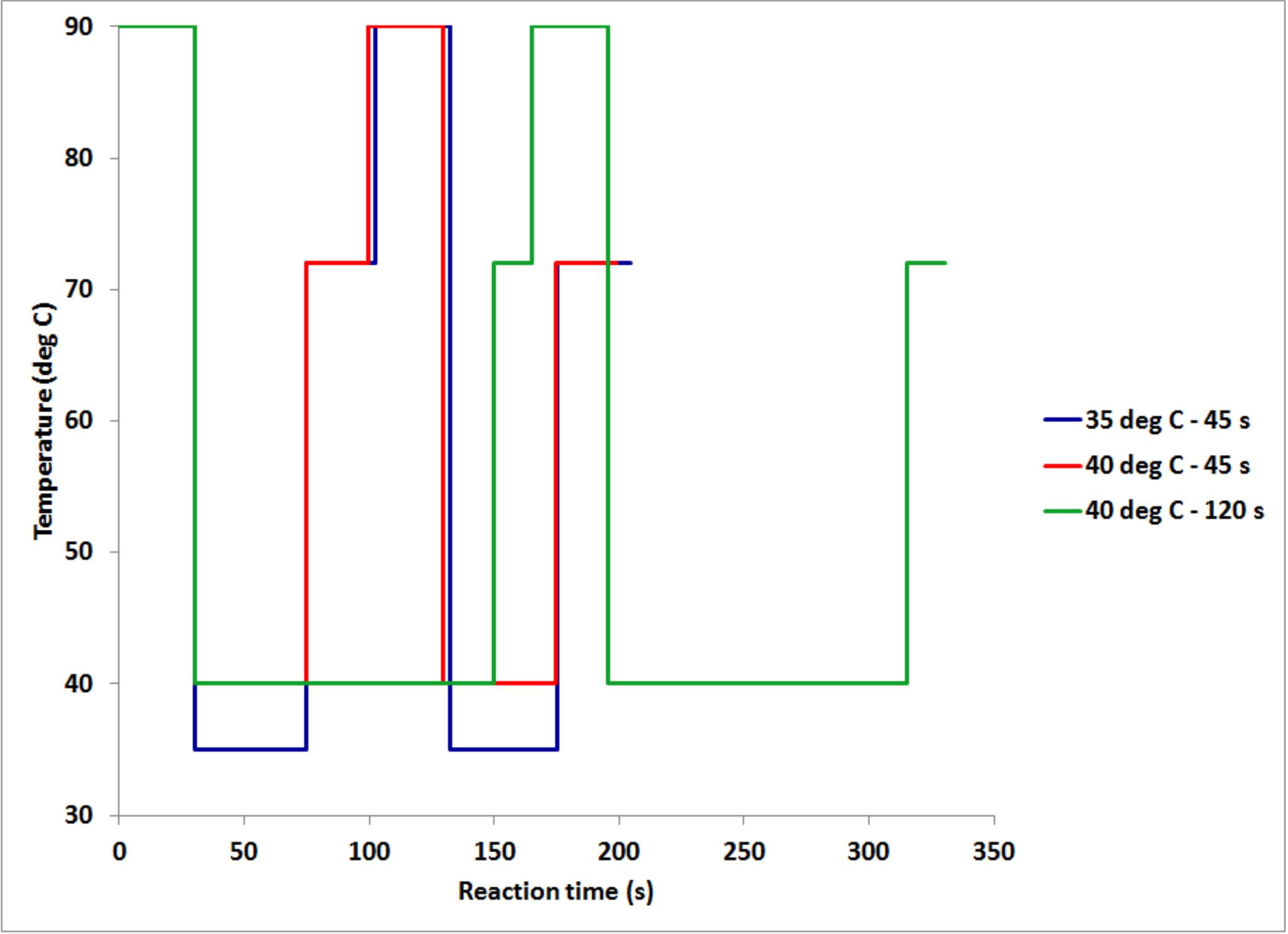}
\label{modified_t_protocol}}
\quad
\subfigure[]{
\includegraphics[width=8cm,height=6cm]{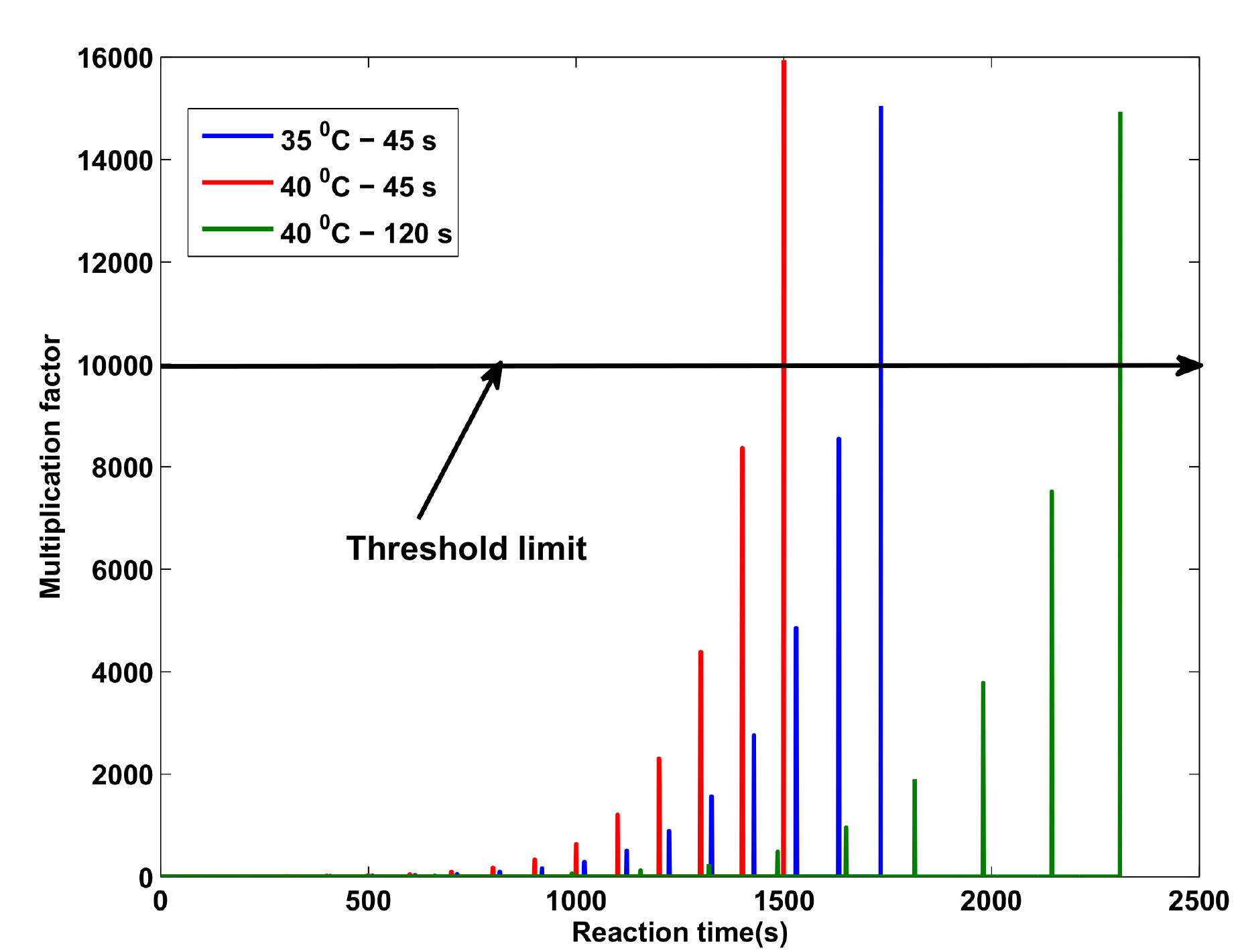}
\label{opt_eta_gg_1}}
\caption{a) Optimized cycling protocol at three different reaction conditions computed based on the cycle switching time criterion depicted in Fig. \ref{optimal_eta}. The legends refer to annealing temperatures and times; b)Geometric growth of DNA for three different conditions }
\end{figure*}

Fig. \ref{modified_t_protocol} illustrates the modified temperature protocol based on this analysis. Fig. \ref{opt_eta_gg_1} compares the geometric growth of DNA with respect to reaction time under the above three reaction conditions.

Thus, we have shown that for a specified control vector trajectory $k(t)$ for a cycle, it is possible to reduce the overall reaction time by application of the time optimal cycle switching criterion (\ref{min_time_opt6}) for geometric growth. In the above case, the optimal choice of annealing temperature is 40 $^0$C. In the above study, however, we did not consider optimization of the annealing step for the minimization of overall reaction time and hence this method naturally optimizes only the extension reaction time at the end of each cycle. This constraint will be relaxed in future work to solve the time optimal control problem (Eq. (\ref{obj_fun_2})) using a TVM model.

\subsubsection{Bilinear time-varying PCR model with and without drift}
In stage 1, the primer and nucleotide concentrations are always much higher than the target DNA concentration and the change in their values is negligible compared to the primer and nucleotide concentrations. Therefore, the primer and nucleotide concentration can be treated as constant and hence the annealing and extension reactions are pseudo-first order. The enzyme binding reaction can also assumed to be a pseudo first order reaction, but not for all the cycles in stage 1. In the last few cycles of this stage, enzyme concentration can be comparable with the target DNA concentration; therefore, except for the last two cycles, it can be assumed that even enzyme binding reaction also is a pseudo first order reaction. Furthermore, it is also possible to assume that the melting reaction is irreversible, as hybridization of ssDNA is negligible due to large excess of primers. Thus, the whole PCR model can be expressed as a linear time varying first order state space system. Since both control and state variables vary with respect to time, to be precise, the PCR model can be expressed as a bilinear control system as follows
\begin{widetext}
\begin{equation}
\frac{dx\left(t\right)}{dt} =\left(A + \sum_{i=1}^{9}B_ik_i\left(t\right)\right)x\left(t\right) \label{time_variant_linear_pcr} \\
\end{equation}
$$k \in \mathbb{R}^{9};~k \geq 0,~~ x,g_i\left(x\right) \in \mathbb{R}^{4n+5},~x \geq 0,~~\Phi x =0$$
where
$$ \Phi =  \left[S_{10}/S_1 - 1, -1 , - c, -1, - c, -1 , S_{20}/S_2 - 1 , -1 , - c, -1 , -c , -1 , -1 \right] $$
$$ c = \left(1 + \frac{\left[N_0\right]}{K_N} \right)$$
\end{widetext}
Since the association of ssDNA is neglected, the corresponding rate constants have been eliminated and therefore, the total number of rate constants is 9 and the total number of states is $4n+5$.
If the whole system is time varying, then $A =0$, $B_i$ is a $4n+5 \times 4n+5$ matrix and
\begin{widetext}
\begin{equation}
B_i x\left(t\right) = g_i \left(x\left(t\right)\right)
\end{equation}
$$x_1 = \left[S_1\right],  x_2 = \left[S_1P_1\right], x_3 =  \left[E.S_1P_1\right], x_{2i+2} = \left[D_i^1\right], x_{2i + 3} = \left[E.D_i^1\right]  \hspace{2mm} \forall i = 1,2,...n-1$$
$$x_{2n+3} = \left[S_2\right],  x_{2n+4} = \left[S_2P_2\right], x_{2n+5} =  \left[E.S_2P_2\right], x_{2i+2n+4} = \left[D_i^1\right], x_{2i + 2n + 5} = \left[E.D_i^1\right]  \hspace{2mm} \forall i = 1,2,...n-1$$
$$ x_{2n + 2} =\left[ E.D_n^1\right], x_{4n + 4} = \left[E.D_n^2\right], x_{4n+5} =\left[DNA\right]$$
\end{widetext}
$$ k = \left[k_m \hspace{2mm} k_1^1 \hspace{2mm}  k_2^1  \hspace{2mm}  k_1^2  \hspace{2mm}  k_2^2 \hspace{2mm} k_1^e \hspace{2mm} k_{-1}^{e} \hspace{2mm} \frac{k_{cat}}{K_N} \hspace{2mm} k_{cat}\rq{} \right]$$
$B_1$ represents the melting reaction. $B_2$ and $B_3$ represent the annealing of the 1st single strand and primer. $B_4$ and $B_5$ represent the annealing of the 2nd single strand and primer. $B_6$ and $B_7$ represent the enzyme binding reactions. $B_8$ and $B_9$ represent the extension reaction. Eq. (\ref{time_variant_linear_pcr}) specifies a TVM model for stage 1.  
\begin{itemize}
\item If the melting and annealing rate constant alone is varied (TVMD for annealing step) then
\begin{equation}
\frac{dx\left(t\right)}{dt} =\left(A + \sum_{i=1}^{5}B_ik_i\left(t\right)\right)x\left(t\right)
\end{equation}
\noindent
Here $A$ is a $4n+5\times4n+5$ matrix and
\begin{equation}
Ax\left(t\right) = f\left(x\left(t\right)\right) = \left(\sum_{i=6}^{9} k_iB_i\right)x\left(t\right)
\end{equation}
\item If the enzyme binding and extension rate constant alone is varied (TVMD for extension step) then
\begin{equation}
\frac{dx\left(t\right)}{dt} =\left(A + \sum_{i=6}^{9}B_ik_i\left(t\right)\right)x\left(t\right)
\end{equation}
\noindent
Here $A$ is a $4n+5\times4n+5$ matrix and
\begin{equation}
Ax\left(t\right) = f\left(x\left(t\right)\right) = \left( \sum_{i=1}^{5} Ak_i \right)x\left(t\right)
\end{equation}
\end{itemize}
For temperature control, constraints analogous to those in Eq. (\ref{arrhenius}) are imposed on the vector of rate constants. The linear structure of the PCR state equations in Stage 1 is convenient for refinement of kinetic parameter estimates via linear filtering \cite{Stengel}, which will be discussed in future work Moreover, it enables analytical solution of the state equations for each step of PCR.

\subsection{Stage 2: Multistep cycling}
This stage is comprised of all the cycles at which the enzyme concentration is comparable to or less than the target DNA concentration.
We have shown in Section \ref{simplex} that in Stage 2, annealing reaction time needs to be increased in order to maximize the DNA concentration, since enzyme must rebind to excess SP molecules after dissociating from fully extended dsDNA. Though it is possible to consume all the ssDNA by fixing a long annealing time, this does not exploit the higher rate of extension at higher temperatures.

\begin{figure*} 
\centering
\subfigure[]{
\includegraphics[width=8cm,height=6cm]{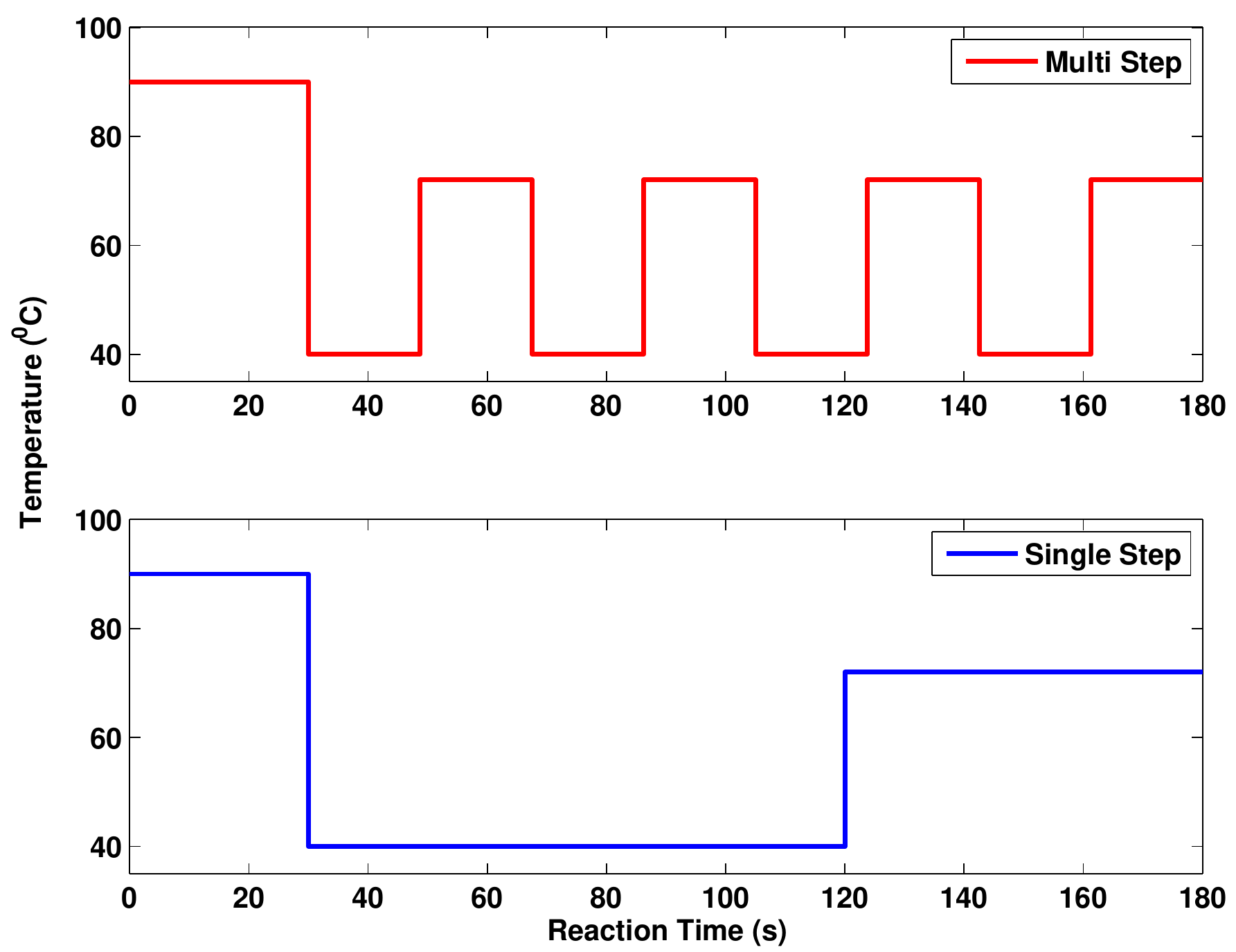}
\label{muti_step_temp}}
\quad
\subfigure[]{ 
\includegraphics[width=8cm,height=6cm]{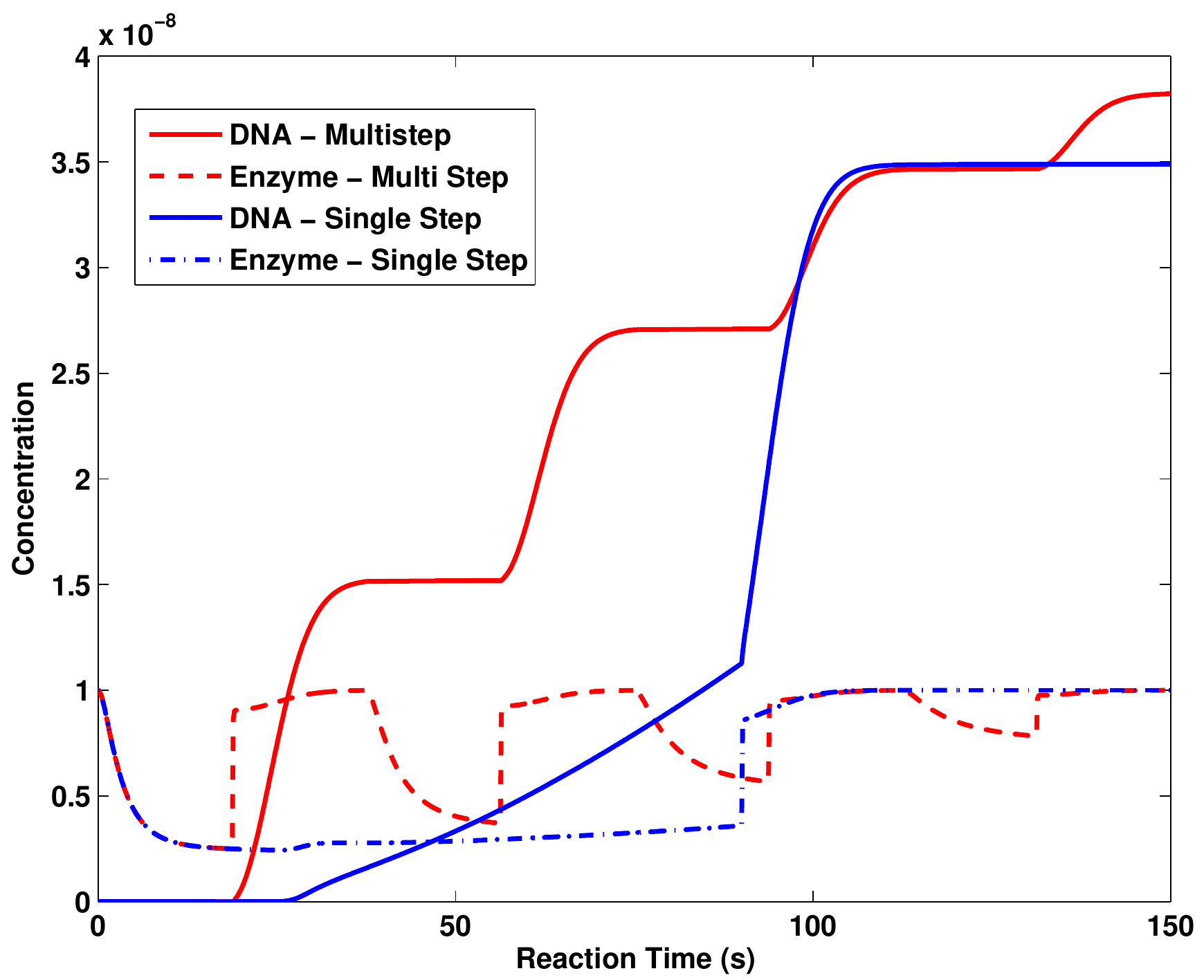}
\label{muti_step}}
\caption{Multistep PCR. a) Temperature profile for multiple annealing step (multi-step) and single annealing step PCR. b) DNA concentration profile in multi-step PCR. In multi-step PCR, in a PCR cycle, annealing and extension was repeated four times. In a single annealing step PCR, a long annealing time was maintained. Enzyme and template concentrations are 10 nM and 20 nM, respectively, the primer concentration is 200 nM and the nucleotide concentration is 800 $\mu$ M.}
\end{figure*}

Therefore, in Stage 2 it is optimal to conduct annealing and extension multiple times in order to obtain higher cyclic efficiency.
If the initial concentration of the target DNA at the beginning of a cycle is $D_0$, which is greater than or equal to the enzyme concentration, then the number of annealing and extension steps that needs to be conducted within a given cycle to double the concentration of $D_0$ is given as
\begin{equation}
N_s = \frac{2D_0}{E_0} \label{no_steps}
\end{equation}
Based on Eq. (\ref{no_steps}), the number of annealing and extension steps required to double the target concentration is $\mathcal{O}(D_0/E_0)$. By using the appropriate number of annealing and extension steps per cycle, we can reduce the overall PCR reaction time required to achieve a specified DNA concentration. Fig. \ref{muti_step} shows the evolution of DNA and enzyme molecules in a single annealing step and multistep PCR for a cycle in which the initial concentration of target and enzyme are 10 and 20 nM, respectively and the goal is to achieve 40 nM DNA concentration. Fig. \ref{muti_step_temp} compares the temperature profiles for  regular and multistep PCR.

For both the fixed time problem (Eq. (\ref{u_manipulated_1})) and minimal time problem (Eq. (\ref{obj_fun_2})), a multi-step strategy is generally a property of the optimal solution.  In each of cycle of this stage, in order to double to concentration of the template DNA, annealing and extension needs to be conducted multiple times. In minimum time optimal control of geometric growth we have shown that 100\% efficiency need not be achieved. In such cases the number of steps in a cycle will be less than or equal to the number of steps that was predicted by Eq. (\ref{no_steps}). In order to determine the optimal number of steps, the minimal time optimal control problem described above needs to be solved, either for a single cycle (if the desired single cycle efficiency is known) or for $m \geq 1$ (for more than 1 cycle). For both the fixed time and minimal time problems, the optimal solutions $k^*(t)$ for stage 2 are not periodic.

\section{Conclusion}
In this work, the dynamics of DNA amplification reactions have been formulated from a control theoretic standpoint. Optimal control problems for maximization of a desired target DNA concentration (Eq. (\ref{u_manipulated_1})) and minimization of the overall cycle time (Eq. (\ref{obj_fun_2})) have been specified and a strategy for the optimal synthesis of the temperature cycling protocol has been presented. Future work will consider the derivation of the optimality conditions and optimal control laws for these problems \cite{chakrabarti2014}, as well as those for control problems pertaining to other DNA amplification objectives -- including those that involve the co-amplification of multiple DNA sequences \cite{li2008replacing} and the automated design of new types of PCR reactions. When applied to given sequences through the use of state-of-the-art dynamic optimization strategies, these laws prescribe the optimally controlled amplification dynamics of DNA.

We have presented a general framework for optimal control of DNA amplification in terms of sequence- and temperature-dependent rate constants. For a fixed sequence, solution of the control problem provides the optimal temperature profile corresponding to the specified amplification objective. However, based on the sequence-dependent kinetic model, it is also possible to consider optimization of replication efficiency through sequence mutations given a specified time-varying temperature profile. 
More generally, the framework presented is capable of accommodating problems wherein replication efficiency is optimized through successive iterations of sequence evolution and changes in the time-varying temperature profile. According to control problem formulation (Eqs. (\ref{u_manipulated_1},\ref{u_manipulated_2})), this relaxes constraints 
on the time-varying rate constants, such that the replication dynamics are no longer controlled by only a single function of time.

The latter generalized dynamic optimization problem is applicable to primer design for optimally controlled DNA amplification. It is also of interest in the context of the chemical evolution of the earliest nucleic acid amplification reactions \cite{Szostak2009}. Recent studies have considered how in the early stages of prebiotic evolution, replication of nucleic acids in minimal protocellular compartments may have occurred nonenzymatically \cite{Szostak2012}, followed by the evolution of ribozyme polymerases (replicases) within these compartments that were capable of self-replication \cite{Holliger2013}. Both stages of chemical evolution 
likely required time-varying environmental inputs in order to promote strand separation and polymerization in an alternating fashion. Indeed, studies have shown that protocell membranes may have been capable of withstanding temperatures of up to nearly 100 $^0$C \cite{Szostak2008}, and that PCR amplification within 
supramolecular vesicles can induce vesicles to divide, thus suggesting that the amplification of nucleic acids through temperature cycling can be coupled to the replication of protocells \cite{Kurihara2011}.
Efforts are underway to explore the extent to which an optimal temperature profile for control of nucleic acid amplification, derived using the methods described herein, can accelerate the evolution of increased replication efficiency of nucleic acids through mutations.

\end{document}